\documentclass[11pt,letterpaper]{article}
\usepackage[rflt]{floatflt}
\usepackage{fullpage}
\usepackage[latin1]{inputenc}
\usepackage{subfig}
\usepackage{graphicx}
\usepackage{epsfig}
\usepackage{color}
\usepackage{enumerate}
\usepackage{verbatim}
\usepackage{amsthm, amssymb}
\usepackage{amsmath}
\usepackage{mdwlist}

\usepackage{algorithm}
\usepackage{algorithmicx}
\usepackage{algpseudocode}
\usepackage{fixltx2e}
\usepackage[all=normal,lists,paragraphs]{savetrees}
\date{}
\MakeRobust{\overrightarrow}

\newtheorem{theorem}{Theorem}
\newtheorem{lemma}[theorem]{Lemma}

\newtheorem{observation}[theorem]{Observation}
\newtheorem{definition}[theorem]{Definition}
\newtheorem{corollary}[theorem]{Corollary}

\def\adj{\operatorname{adj}}

\def\term{\operatorname{term}}
\newcommand{\lca}{\ensuremath{\mathit{lca}}}

\def\dist{\mathit{dist}}

\newcommand{\out}[1]{}

\DeclareMathOperator{\sgn}{sgn}

\title{\bf Algorithmic Applications of Baur-Strassen's Theorem: Shortest Cycles, Diameter and Matchings}
\author{
Marek Cygan \thanks{Institute of Informatics,
University of Warsaw {\tt cygan@mimuw.edu.pl}.}
\and
Harold N. Gabow \thanks{Department of Computer Science, University of Colorado at Boulder {\tt
hal@cs.colorado.edu}.}
\and
Piotr Sankowski\thanks{Institute of Informatics,
University of Warsaw and Department of Computer and System Science,
Sapienza University of Rome {\tt sank@mimuw.edu.pl}.}
}

\begin{document}

\begin{titlepage}
\def\thepage{}
\thispagestyle{empty}
\maketitle

\begin{abstract}
Consider a directed or an undirected graph with integral edge weights from the set $[-W, W]$, that does not contain
negative weight cycles. In this paper, we introduce a general framework for solving problems
on such graphs using matrix multiplication. The framework is based on the usage of Baur-Strassen's theorem and of
Strojohann's determinant algorithm. It allows us to give new and simple solutions to the following problems:
\begin{description}
\item[Finding Shortest Cycles] -- We give a simple $\tilde{O}(Wn^{\omega})$ time algorithm for finding shortest cycles in undirected and
directed graphs. For directed graphs (and undirected graphs with non-negative weights) this matches the time bounds obtained in 2011 by Roditty and Vassilevska-Williams. On the other hand,
no algorithm working in $\tilde{O}(Wn^{\omega})$ time was previously known for undirected graphs with negative weights.
Furthermore our algorithm for a given directed or undirected graph detects whether it contains a negative 
weight cycle within the same running time.
\item[Computing Diameter and Radius] -- We give a simple $\tilde{O}(Wn^{\omega})$ time algorithm for computing 
a diameter and radius of an undirected or directed graphs. To the best of our knowledge 
no algorithm with this running time was known for undirected graphs with negative weights.
\item[Finding Minimum Weight Perfect Matchings] -- We present an $\tilde{O}(Wn^{\omega})$ time algorithm for finding minimum weight perfect matchings in undirected graphs. This resolves an open problem posted by Sankowski in 2006, who presented such an algorithm but only in the case of bipartite graphs.
\end{description}
We believe that the presented framework can find applications for solving larger spectra of related problems. As an illustrative
example we apply it to the problem of computing a set of vertices that lie on cycles of length at most $t$, for some $t$.
We give a simple $\tilde{O}(Wn^{\omega})$ time algorithm for this problem that improves over the $\tilde{O}(Wn^{\omega}t)$ 
time algorithm given by Yuster in 2011.

In order to solve minimum weight perfect matching problem we develop a novel combinatorial interpretation of the dual solution
which sheds new light on this problem. Such a combinatorial interpretation was not know previously, and is of independent interest.
\end{abstract}
\end{titlepage}
\section{Introduction}
The application of matrix multiplication to  graph problems has been  actively studied  in  recent years. The 
special case of 
unweighted graphs  is well understood. For example, $\tilde{O}(n^{\omega})$ time algorithms for  finding
shortest cycles~\cite{Itai77} have been known for 35 years\footnote{The $\tilde{O}$ notation ignores factors of
$\log n$ and $\log W$.
$\tilde O(n^{\omega})$ is the time needed for a straight-line program to multiply two $n\times n$ matrices; $\omega$ is called matrix multiplication exponent.}. But similar results for weighted graphs were obtained only 
last year, by Roditty and Vassilevska-Williams~\cite{RodittyW11}. Their algorithm works in $\tilde{O}(Wn^{\omega})$ time, where $W$ is the largest magnitude of an edge weight. Two similar problems 
on weighted graphs where there has been considerable effort, 
but the full answer has not been achieved, are
diameter and perfect weighted matching. This paper
introduces a general framework that gives simple%
\footnote{An objective sense in which our algorithms are simple is their use of algebra: 
The power of our algebraic algorithms comes from black-box routines, and the algorithms themselves
use only elementary algebraic ideas.}
solutions to all three of these problems and others. In the following unless otherwise
stated, we work with graphs that contain edges with possibly negative weights but no negative  cycles. 
We obtain the following results.

\paragraph{Finding Shortest Cycles} We give a simple $\tilde{O}(Wn^{\omega})$ time algorithms for finding shortest cycles in undirected and
directed graphs. In the case of directed graphs the algorithm reduces the problem to one determinant computation for a polynomial matrix. On the
other hand, the undirected case requires handling short 2-edge cycles in a proper way. The idea used here is an extension of the algorithm by Sankowski contained in~\cite{Sankowski05}, that allowed to test whether a graph contains negative weight cycle. For directed graphs (and undirected graphs with non-negative weights), our bounds match the ones obtained in 2011 by Roditty and Vassilevska-Williams~\cite{RodittyW11}, whereas for undirected graphs with negative weights no $\tilde{O}(Wn^{\omega})$ time algorithm
was previously know for this problem. For all related results see Table~1 and Table~2. For undirected graphs with negative
weights the problem reduces to $n$ computations of minimum weight perfect matchings~\cite{Edmonds67}.

Furthermore our algorithm for a given directed or undirected graph detects whether it contains a negative 
weight cycle within the same running time.
\paragraph{Computing Diameter and Radius} We present a  simple $\tilde{O}(Wn^{\omega})$ time algorithm for computing a diameter and radius of undirected and
directed graphs. This algorithm combines determinant computations with binary search.
Since computing all pairs shortest paths suffices to find both diameter and radius $\tilde{O}(Wn^{\omega})$ time algorithm follows from~\cite{Shoshan99}
in case of undirected graphs with non-negative weights.
Moreover by generalizing the ideas of~\cite{Shoshan99} used for non-negative weights by applying random sampling one can obtain the same running time for directed
graphs without negative weight cycles.
However, to the best of our knowledge, all previous solutions to this problem in undirected graphs with negative weights reduced 
the problem to $n$ computations of minimum weight perfect matchings.
For other related results see Table~3 and Table~2. 


\paragraph{Finding Minimum Weight Perfect Matchings} We present an $\tilde{O}(Wn^{\omega})$ time algorithm for finding minimum weight perfect matchings
in undirected graphs. This resolves an open problem posted by Sankowski in 2006~\cite{Sankowski09}, who presented such an algorithm but only in the case of bipartite graphs. Some advance on this problem has been recently given by Huang and Kavitha~\cite{HuangK12}, who have shown an $\tilde{O}(Wn^{\omega})$ time algorithm for the maximum weight matching problem. However, the weighted perfect matching problem is more
involved and no reduction similar to the one presented in~\cite{HuangK12} is known to work. Previously, similar reduction was given in~\cite{kao-99} for maximum weight bipartite matching. Nevertheless, to solve minimum weight perfect matching problem even in bipartite graphs one is bound to use more structured techniques. Actually, we need to develop such a technique for general graphs. We give a novel combinatorial interpretation of the
dual problem. Such an interpretation for general matching problem was not know previously and is of significant independent interest.
For the summary of different algorithms see Table~2.
\vspace{0.4cm}

\noindent
\begin{tabular}{cc}
  \begin{minipage}{0.48\textwidth}
    {\scriptsize\noindent
      \begin{tabular}{|p{3.8cm}|p{3.2cm}|} \hline
        Complexity & Author \\ \hline
        $O(nm+n^2\log n)$ \newline directed & Johnson (1977) \cite{johnson}\\ \hline
        $O(n^{\omega})$ nonnegative unweighted & Itai and Rodeh (1977) \cite{Itai77} \\ \hline
        $O(W^{0.681}n^{2.575})$ directed & Zwick (2000) \cite{zwick}\\ \hline
        $O(nm+n^2\log\log n)$ directed & Pettie (2004) \cite{pettie}\\ \hline
        $O(n^3 \log^3 \log n/ \log^2 n)$ directed & Chan (2007) \cite{Chan2007}\\ \hline
        $\tilde{O}(Wn^{\omega})$ directed and nonnegative undirected & Roditty and Vassilevska-Williams (2011) \cite{RodittyW11}\\ \hline
        $\tilde{O}(Wn^{\omega})$ & this paper\\ \hline
      \end{tabular}
      \begin{minipage}{\textwidth}
      \vspace{0.2cm}
      {\bf Table 1: The complexity results for the shortest cycle problem. }
      \end{minipage}
    }
  \end{minipage}
  &
  \begin{minipage}{0.48\textwidth}
    {\scriptsize\noindent
      \begin{tabular}{|p{4.1cm}|p{2.9cm}|} \hline
        Complexity & Author \\ \hline
        $O(n^2m)$ & Edmonds (1965) \cite{e} \\ \hline
        $O(n^3)$ & Lawler (1973) \cite{lawler-73} and Gabow (1974) \cite{gabow-76}  \\ \hline
        $O(nm \log n)$ & Galil, Micali and Gabow (1982) \cite{galil-86}  \\ \hline
        $O(n(m \log \log \log_{\frac{m}{n}} n + n \log n))$ & Gabow,
        Galil and Spencer (1984) \cite{gabow-89} \\ \hline
        $O(n^{\frac{3}{4}} m \log W)$ & Gabow (1985) \cite{gabow-85-2}  \\ \hline
        $O(n(m+n \log n))$ & Gabow (1990) \cite{gabow-90}  \\ \hline
        $O(\sqrt{n \alpha(m,n)\log n}\ m \log (nW))$ & Gabow and Tarjan (1991) \cite{gt91}  \\ \hline
        $\tilde{O}(Wn^{\omega})$ & this paper \\ \hline
      \end{tabular}
      \begin{minipage}{\textwidth}
      \vspace{0.2cm}
      {\bf Table 2: The complexity results for the minimum weight perfect matching problem.
      Note that first, third, fourth and fifth results solve the more general All Pairs
      Shortest Paths problem.}
      \end{minipage}
    }
    \vspace{-0.45cm}
  \end{minipage}\\
  \begin{minipage}{0.48\textwidth}
  \vspace{0.2cm}
    {\scriptsize\noindent
      \begin{tabular}{|p{4cm}|p{3.0cm}|} \hline
        Complexity & Author \\ \hline
        $O(nm+n^2\log n)$ directed & Johnson (1977) \cite{johnson}\\ \hline
        $O(n\min(m\log n,n^2))$ undirected & Gabow (1983) \cite{hal83} \\ \hline
        $\tilde{O}(Wn^{\omega})$ nonnegative undirected &  Shoshan and Zwick (1999) \cite{Shoshan99} \\ \hline
        $O(W^{0.681}n^{2.575})$ directed & Zwick (2000) \cite{zwick}\\ \hline
        $O(n^3 \log^3 \log n/ \log^2 n)$ \newline directed & Chan (2007) \cite{Chan2007}\\ \hline
        $O(Wn^{\omega})$ directed & folklore+\cite{Shoshan99}+\cite{zwick}\\ \hline
        $\tilde{O}(Wn^{\omega})$ & this paper \\ \hline
      \end{tabular}
      \begin{minipage}{\textwidth}
      \vspace{0.2cm}
      {\bf Table 3: The complexity results for the diameter and radius problem. Note that the first five results
       solve the more general All Pairs Shortest Paths problem.}
      \end{minipage}
    }
        \end{minipage}
  &
  \begin{minipage}{0.48\textwidth}
    {\scriptsize\noindent
      \begin{tabular}{|p{3.0cm}|p{4.0cm}|} \hline
        Complexity & Author \\ \hline
        $\tilde{O}(nm)$ & Suurballe and Tarjan (1984) \cite{suuraballe} \\ \hline
        $\tilde{O}(t Wn^{\omega})$ undirected & Yuster (2011) \cite{yuster11}\\ \hline
        $\tilde{O}(t^{4-\omega} Wn^{\omega})$ directed & Yuster (2011) \cite{yuster11}\\ \hline
        $\tilde{O}(Wn^{\omega})$ & this paper\\ \hline
      \end{tabular}
      \begin{minipage}{\textwidth}
      \vspace{0.2cm}
      {\bf Table 4: The complexity results for the problem of finding vertices that lie on cycles of length at most $t$.}
      \end{minipage}
      }
  \end{minipage}
\end{tabular}

\subsection{The Framework}
Our framework is based on two seminal results. First of all we use Storjohann's algorithm~\cite{storjohann03} that computes the determinant of a degree $d$ polynomial matrix in $\tilde{O}(dn^{\omega})$ time. All the above graph problems can be encoded as a determinant problem on a
polynomial matrix. However the determinant is just one number and does not provide enough information to decode the whole solution. Here
we use the Baur-Strassen Theorem~\cite{Baur1983}, which shows how to compute all
partial derivatives of a function in the same asymptotic time as computing the function itself. For a simple constructive proof please check~\cite{Morgenstern85}. This theorem allows us to magnify the output of the algorithm from $1$ number to $n^2$ numbers. The
algorithms  obtained in this way are very simple and work in three phases: compute the determinant of an appropriately defined matrix; apply Baur-Strassen 
to the result; decode the output. Even for minimum weight matching our algorithm is simple, and computes the dual solution in just a few lines of pseudocode. Based on these examples, we believe
this framework will find application to a wider spectrum of problems. The full version of the paper will contain a more extensive review of possible applications. Here we give one more illustration:
a simple $\tilde{O}(Wn^{\omega})$ time algorithm that 
finds every vertex that lies on  a cycle of length  $\le t$, for a given arbitrary $t$ (for a directed or undirected
weighted  graph, with negative edges allowed but no negative cycles). This improves the algorithms proposed by Yuster in 2011~\cite{yuster11} (see Table~4; those algorithms do not allow negative edges). 

The paper is organized as follows. In the next two sections we introduce needed tools and give the main definitions. In Section~\ref{section-cycles} we introduce our framework and give algorithms for the shortest cycle problem in directed graphs.
This motivates the introduction of the framework in Section~\ref{section-meta}. Section~\ref{section-matchings} contains the algorithm for minimum weight perfect matching problem. The next section presents the application of our framework for computing the diameter of a graph.
Section~\ref{section-cycles-u} introduces the ideas needed to compute shortest cycles in undirected graphs with positive weights. In Section~\ref{section-undirected-negative} we join the ideas from all previous section to solve the shortest cycle problem and the diameter problem
in undirected graphs with negative weights. Finally, in Section~\ref{section-cycles-passing} we apply our framework to find the set of vertices that lie on a cycle of lenght at most $t$.

\section{Preliminaries}

Our approach is based on three main ingredients.

\paragraph{Linear Algebra Algorithms}
\label{subsection-linear-algebra-algorithms}
Storjohann~\cite{storjohann03} has made an important addition to 
 the set of problems solvable in $O(n^{\omega})$ arithmetic operations:
the determinant and the rational system solution for polynomial matrices.

\begin{theorem}[Storjohann '03]
  \label{theorem-storjohann}
Let $K$ be an arbitrary field,
  $A \in K[y]^{n \times n}$ a polynomial matrix of degree $d$,
  and $b \in K[y]^{n \times 1}$ a polynomial vector of the same degree. Then
  \begin{itemize}
  \item rational system solution $A^{-1}b$ (Algorithm~5 \cite{storjohann03}),
  \item determinant $\det(A)$ (Algorithm~12 \cite{storjohann03}),
  \end{itemize}
  can be computed in $\tilde{O}(dn^{\omega})$ operations in $K$, with
  high probability.
\end{theorem}

For the next section, note that both algorithms of Storjohann can be written
as straight-line programs. The randomization does not pose a problem as it is just
used at the very beginning to generate a polynomial of degree $d$ that does not
divide $\det(A)$. After this the algorithms are deterministic. Our applications will 
work over a finite field (see Section~\ref{section-zippel-schwartz}) so there is no risk
of manipulating huge integers. Finally usage of the FFT to perform multiplication of
degree $nd$ polynomials does not pose a problem.

\paragraph{Baur-Strassen Theorem}
Another astonishing result is the Baur-Strassen Theorem
from 1983~\cite{Baur1983,Morgenstern85}. 
It was used to show that matrix multiplication is no harder than determinant computation
when considering algorithms that can be written as straight-line programs.
Such a statement
is unexpected when you realize that matrix multiplication returns
$n^2$ numbers and determinant computation just one.
However it is possible to increase the number of outputs by modifying the algorithm appropriately.
Let $T(f_1,\ldots,f_k)$ denote the time needed to compute functions $f_1,\ldots, f_k$, all at the same given point.

\begin{theorem}[Baur-Strassen '83]
\label{theorem-baur-strassen}
For straight-line programs computing $f(x_1,\ldots, x_n)$,
\[
T(f,\frac{\partial f}{\partial x_1},\ldots, \frac{\partial f}{\partial x_n})\le 5T(f).
\]
\end{theorem}
Thus we can compute all partial derivatives of a function $f$ in the same
asymptotic time as is used to compute $f$. \cite{Morgenstern85}
shows a RAM routine to compute all $n$ partials 
can be constructed in
time $O(T(f))$
as well.

\paragraph{Schwartz-Zippel Lemma}
\label{section-zippel-schwartz}
Our approach is to
 encode a graph problem in a symbolic matrix whose
determinant is (symbolically) non-zero if and only if
the problem has a solution. 
The Schwartz-Zippel Lemma~\cite{z79,s80} provides the non-zero test:

\begin{corollary}
  \label{corollary-zippel-schwartz}
For any prime $p$,
  if a (non-zero)
  multivariate polynomial of degree $d$ over $Z_p$ is evaluated at a random point,
  the probability of false zero is 
$\le d/p$.
\end{corollary}

We will choose primes $p$ of size $\Theta(n^c)$ for some constant $c$.
In a RAM machine with word size $\Theta(\log n)$, arithmetic modulo $p$ can be realized in constant time.
Moreover we would like to note that multivariate determinant expressions
have been used several times in previous work,
including Kirchhoff's Matrix-Tree Theorem,
results of Tutte and Edmonds on perfect matchings,
and recent result of Bj{\"o}rklund~\cite{bjorklund} for undirected Hamiltonicity.

\section{Definitions: Graphs with Integral Weights}
\label{section-definitions}
A {\it weighted $n$-vertex graph} $G$ is a tuple $G=(V,E,w,W)$, where
the vertex set is given by $V = \{1, \ldots, n\}$, $E \subseteq V
\times V$ denotes the edge set, and the function $w:E \to
[-W, W]$ ascribes weights to the edges. In
this paper we consider both undirected and directed graphs. Hence,
$E$ might denote either a set of unordered pairs for undirected graphs or
a set of ordered pairs for directed graphs. We define {\it a weight
of the edge set} $F\subseteq E$ to be $e(F) = \sum_{e\in F}w(e)$.

Consider a path $p = v_1, v_2, \ldots, v_k$ of length $k$. The {\it weight}
of this path is simply equal to the weight of the edge set of $p$ and denoted by by $w(p)$.
The {\it distance} from $v$ to $u$ in $G$, denoted by $\dist_G(v,u)$, is
equal to the minimum weight of the paths starting at $v$ and ending in
$w$. A path from $v$ to $u$ with minimum weight is called {\it a shortest path}.
If for a path $p$ we have $v_1 = v_k$ then the path is called {\it a cycle}.

In the {\it shortest cycle problem} we are given a weighted graph $G$
and we need to compute the shortest (i.e., minimum weight) cycle in $G$. In this problem we
assume that the graph contains no negative weight cycles. There is a
subtle difference between the directed and undirected versions of the problem.
The standard approach to reduce the undirected problem to directed problem by bidirecting the edges
does not work. The resulting graph can contain cycles that pass though the same edge
twice in both directions. Such cycles do not exist in the undirected graph.
Moreover, when the undirected graph contains negative edges the resulting directed graph
contains negative weight cycles, even if the undirected graph did not. 

In the {\it diameter problem} we are given a weighted graph $G$ (directed or
undirected) and are asked to find the a pair of vertices $v,u\in V$ that maximizes
$\dist_G(v,u)$. Note that here the reduction of undirected problem to directed
one by bidirecting the edges works on the in the case when the edges are positive. 

A {\it matching} $M$ in $G$ is a set of edges such that each vertex is incident at most one edge from $M$.
In a {\it perfect matching} each vertex is incident to exactly one matched edge.
A {\it minimum weight perfect matching} is a perfect
matching $M$ in a weighted undirected graph  that minimizes the total weight
$w(M) = \sum_{e\in M} w(e)$. Many other notions of "optimum weighted matching"
reduce to minimum weight perfect matching:
 A {\it maximum weight perfect matching} is equivalent to a
minimum weight perfect matching for weights $w'(e):=-w(e)$. 
A {\em minimum weight cardinality $k$ matching} (i.e., exactly $k$ edges are to be matched)
is a minimum weight perfect matching on the graph with $n-2k$ articial vertices, each
joined to every original vertex by a zero-weight edge. 
A {\it maximum weight matching} (i.e., we want to maximize the total weight of the matched edges)
 is a maximum weight perfect matching
on the graph that has one artificial vertex if $n$ is  odd, plus new 
zero-weight edges that make the 
graph complete. There is a reduction going in the other direction:
we set edge weights to $w'(e):=nW+w(e)$. However we are interested in time bounds that
are linear in $W$, so this reduction is not of interest.

%
%

\section{Shortest Cycles in Directed Graphs}
\label{section-cycles}
Let $K$ be an arbitrary field. A {\it symbolic polynomial} $\tilde{p}[y]$ is a multivariate polynomial over a set of variables $y \cup X$ over $K$.
We denote the set of symbolic polynomials by $\tilde{K}[y] = K[y \cup X]$. We write as well $\tilde{K}$ to
denote the set of multivariate polynomials $K[X]$.
A {\it symbolic matrix polynomial} $\tilde{A}[y]\in \tilde{K}[y]^{n \times n}$ is an $n \times n$ matrix
whose entries are symbolic polynomials from $\tilde{K}[y]$.

We shall use a straight-line program that evaluates $\det(\tilde{A}[y])$ 
($\tilde{A}[y]\in \tilde{K}[y]^{n \times n}$)
using Storjohann\rq{}s algorithm. 
Here the goal is to evaluate the  determinant
to a polynomial in one variable $y$.
This program is easily
constructed: 
Start with the original straight-line program that evaluates $\det(A)$ ($A \in K[y]^{n \times n}$) using  Storjohann\rq{}s algorithm.
Prepend assignment statements of the form $a_{ijk}\leftarrow \tilde a_{ijk}$, where
$a_{ijk}$ is the variable in the original program for the coefficient  of $y^k$ in the entry $A_{ij}$,
and  the corresponding coefficient in $\tilde{A}[y]$ is  $\tilde a_{ijk}\in \tilde K$. 
In our applications each of 
these new assignment statements  uses $O(1)$ time, so the extra time can be ignored. Also note
 that all arithmetic in our
straight-line programs is done in $K=Z_p$ for a chosen prime $p$.

We say that $\sigma:X \to K$
is an {\it evaluation function}. We define define $\tilde{p}[y]|_{\sigma}$ to be a one variable polynomial in $y$ with all variables $x \in X$ substituted by $\sigma(x)$.

Let us define a \emph{symbolic adjacency matrix} of the weighted directed
graph \mbox{$\overrightarrow{G}=(V,E,w,W)$} to be the symbolic matrix polynomial $\tilde{A}(\overrightarrow{G})$ such
that
\[
\tilde{A}(\overrightarrow{G})_{i,j} =
 \left\{
 \begin{array}{rl}
 x_{i,j}y^{w((i,j))} & \textrm{if  } (i,j) \in E, \\
 0 & \textrm{otherwise,}
\end{array}
\right.
\]
where $x_{i,j}$ are unique variables corresponding to the edges of
$\overrightarrow{G}$. Hence, $X$ is the set of all variables $x_{i,j}$.
For a multi-variable polynomial $q$, let us denote by:
\begin{itemize}
\item $\deg^*_y(q)$ -- the degree of the smallest degree term of $y$ in $q$,
\item $\term^d_y(q)$ -- the coefficient of $y^d$ in $q$,
\item $\term^*_y(q)$ -- $\term^d_y(q)$ for $d=\deg^*_y(q)$.
\end{itemize}

If
$q$ is the 0 polynomial, $\deg^*_y(q):=\infty$.
In the following we assume that $K := Z_p$, i.e., we work over a finite field of order $p$ for some prime number $p$.

We say that a nonempty set of disjoint cycles $\mathcal{C}$ in $\overrightarrow{G}$ is {\it a cycle packing}
if every vertex of $\overrightarrow{G}$ belongs to at most one cycle in $\mathcal{C}$. Observe that
a minimum weight cycle packing is either a shortest cycle or a collection of shortest cycles all of weight $0$.

\begin{lemma}
\label{lemmaDeterminant}
  Let $\overrightarrow{G}$ be a directed weighted graph. 
The minimum weight
  of a cycle packing in $\overrightarrow{G}$
equals $\deg^*_y(\det(\tilde{A}(\overrightarrow{G})+I)-1)$. 
This weight is also the weight of a shortest cycle if $G$ has no negative cycle. 
Finally for any $G$ and any $p\ge 2$, all non-zero terms in $\det(\tilde{A}(\overrightarrow{G})+I)$ are
  non-zero over a finite field $Z_p$.
\end{lemma}

\begin{proof}
  By definition 
  \begin{equation}
    \label{equationPaths2}
    \det\left(\tilde{A}(\overrightarrow{G})+I\right) =
    \sum_{p \in \Gamma_n} \sigma(p) \prod_{k=1}^{n}
    \left(\tilde{A}(\overrightarrow{G})+I\right)_{k,p_k},
  \end{equation}
  where $\Gamma_n$ is the set of $n$-element permutations, and $\sigma(p)$ is
  the sign of the permutation $p$. A permutation $p$ defines a set of directed edges $\mathcal{C}_p =
  \{(i,p_i) : 1 \le i \le n\}$. This edge
  set corresponds to a set of cycles given by the cycles of $p$.
  The edge set $\mathcal{C}_p$ includes self-loops for all $i$ such
  that $i=p_i$. Note that $p$ corresponds to a non-zero term
  in the determinant if and only if $\mathcal{C}_p$ contains only edges from $E$
  or self-loops. Hence, after throwing away self-loops $p$ can be identified with
  a cycle packing in $G$, or $\emptyset$.

  Let us now compute the degree of $y$ in the term of the determinant given by $p$.
  This term is obtained by multiplying the elements of the matrix $\tilde{A}(\overrightarrow{G})+I$
  corresponding to the edges of $\mathcal{C}_p$. The degree of
  the term equals the sum of the degrees in its individual
  elements. These degrees encode the weights of the edges or are zero for self-loops. So the
  degree of the term is  the total weight of the
  cycles in $\mathcal{C}_p$, where self-loops have weight zero.
 
The term corresponding to a set of self-loops is equal to $1$.
  Hence, $\det(\tilde{A}(\overrightarrow{G})+I)-1$ contains only terms that correspond  to nonempty
  sets of cycles. So we have proved the first assertion, i.e., the smallest degree of $y$
is the smallest weight of a
packings of cycles.

 For the third assertion note that each monomial in (\ref{equationPaths2}) has coefficient
  $\pm 1$ as each of them contains different variables. Hence each non-zero term 
  is also non-zero over $\mathcal{Z}_p$ for any $p \ge 2$.

  Finally for the second assertion assume $\overrightarrow{G}$ has no  negative cycle.
Let $C$ be the shortest cycle in $\overrightarrow{G}$. It  corresponds to a cycle packing
of weight 
$w(C)$. (The packing
has a self-loop
  for every $v \not\in C$.) In fact this is the minimum weight cycle packing.
In proof take
any  cycle packing. If it contains more than one 
 non-loop cycle discard
all but one  of them to get a packing of no greater weight (since every cycle is nonnegative). By definition
this weight is $\ge w(C)$.
\end{proof}

\paragraph{Computing the Weight of the Shortest Cycle}
\label{section-weight-cycle}
Next 
we apply the ideas of the previous section to 
compute the weight of a shortest cycle. Since $\tilde{A}(\overrightarrow{G})$ may have entries with
negative powers of $y$, we cannot use Storjohann's result to compute its determinant. However
we can multiply it by $y^W$ to make exponents nonnegative and use this identity:
\[\deg^*_y(\det(\tilde{A}(\overrightarrow{G})+I)-1)=
\deg^*_y(\det((\tilde{A}(\overrightarrow{G})+I) y^W)-y^{nW}) - n W .
\]
Combining this idea with the Schwartz-Zippel  lemma we obtain the following algorithm.

\begin{algorithm}[H]
\begin{algorithmic}[1]
\State{Generate a random substitution $\sigma:X \to Z_p$ for a prime $p$ of order $\Theta(n^2)$.}
\State{Compute $\delta= \det((\tilde{A}(\overrightarrow{G})|_{\sigma}+I)y^W)-y^{nW}$ using Storjohann's theorem.}
\State{Return $\deg^*_y(\delta)-nW$.}
\end{algorithmic}
\caption{Computes the weight of the shortest cycle in a directed graph $\overrightarrow{G}$.}
\label{algorithm-cycles}
\end{algorithm}

This algorithm implies.

\begin{theorem}
  \label{theorem-shortest-cycle-weight}
  Let $\overrightarrow{G} = (V,E,w,W)$ be a weighted directed graph without negative weight cycles.
  The weight of the shortest cycle in $\overrightarrow{G}$ can be computed in $\tilde{O}(W
  n^{\omega})$ time, with high probability.
\end{theorem}

Note the algorithm also detects the existence of a negative cycle-- it corresponds to $\delta<0$.

\paragraph{Finding a Shortest Cycle}
\label{section-finding-shortest-cycle}
The above algorithms does not seem to give the cycle by itself. There is, however, a
straightforward way of doing it using the Baur-Strassen theorem.
We say that edge $e$ is {\it allowed} if and only if it belongs to some shortest cycle $C$ in $\overrightarrow{G}$.

\begin{lemma}
\label{lemma-allowed}
The edge $(u,v) \in E$ is allowed if and only if:
$\frac{\partial}{\partial x_{u,v}} \term^*_y(\det(\tilde{A}(\overrightarrow{G})+I)-1)$ is non-zero.
Moreover, for $p\ge 2$, it is non-zero over a finite field $Z_p$.
\end{lemma}
\begin{proof}
The above follows directly by the proof of Lemma~\ref{lemmaDeterminant}. The partial derivative is non-zero if and only if
the variable $x_{u,v}$ exists in some smallest degree term, and so the corresponding edge $(u,v)$ has to lie on some shortest cycle.

%
\end{proof}

Rewrite the expression of the lemma to eliminate negative powers of $y$:
\[\frac{\partial}{\partial x_{u,v}} \term^*_y(\det(\tilde{A}(\overrightarrow{G})+I)-1)=
\frac{\partial}{\partial x_{u,v}} \term^*_y(\det((\tilde{A}(\overrightarrow{G})+I) y^W)-y^{nW}).
\]
This implies the following algorithm and theorem.

\begin{algorithm}[H]
\begin{algorithmic}[1]
\State{Let $\partial_x f$ be the routine given by
the Baur-Strassen Theorem to compute the matrix  of partial derivatives
$\frac{\partial}{\partial x_{u,v}} \term^d_y\left[\det((\tilde{A}(\overrightarrow{G})+I) y^W)-y^{nW}\right]$,
 where $d$ is the degree $\deg^*_y(\delta)$ computed in Algorithm 1. }
\State{Generate a random substitution $\sigma:X \to Z_p$ for a prime $p$ of order $\Theta(n^2)$.}
\State{Compute the matrix $\delta=\partial_x f|_{\sigma}$.}
\State{Take any edge $(u,v)$ such that $\delta_{u,v}\neq 0$.}
\State{Compute the shortest path $p_{v,u}$ from $v$ to $u$ in $\overrightarrow{G}$ using~\cite{Sankowski05,YusterZ05}.}
\State{Return the cycle formed by $(u,v)$ and $p_{v,u}$.}
\end{algorithmic}
\caption{Computes a shortest cycle in a directed graph $\overrightarrow{G}$.}
\label{algorithm-cycles-2}
\end{algorithm}

Note that in Step 1 we are applying the Baur-Strassen Theorem to the straight-line program
for 
$\term^d_y\left[\det((\tilde{A}(\overrightarrow{G})+I) y^W)-y^{nW}\right]$. The latter
is constructed using Storjohann\rq{}s Theorem and the modification described
at the start of this section.

The shortest path computation using algorithms from~\cite{Sankowski05,YusterZ05} takes $\tilde{O}(W
  n^{\omega})$ time and succeeds with high probability, so we obtain:
\begin{theorem}
  \label{theorem-cycles}
  Let $\overrightarrow{G} = (V,E,w,W)$ be a weighted directed graph without negative weight cycles.
  The shortest cycle in $\overrightarrow{G}$ can be found in $\tilde{O}(W
  n^{\omega})$ time, with high probability.
\end{theorem}

\section{Performing Symbolic Computations}
\label{section-meta}
This section gives a formal description of the algebraic operations used in our algorithms. 
Let $\tilde{A}[y]$ be an $n \times n$ symbolic matrix and let $\sigma:X \to Z_p$ be an arbitrary evaluation function.
Moreover
for a prime $p$ 
let $f:\tilde{Z_p}[y]^{n \times n} \to \tilde{Z_p}$ be a symbolic matrix function.

\begin{theorem}
\label{theorem-meta-1}
If the result of $f$ has degree bounded by some polynomial $d_f(n)$ and
$f(\tilde{A}[y]|_{\sigma})$ is computable in $t_f(n)$ time, then
there exists an algorithm that in $O(t_f(n))$ time
checks whether $f(\tilde{A}[y])$ is symbolically non-zero over $Z_p$ 
with error probability $\le d_f(n)/p$.
\end{theorem}

\begin{proof}
  The algorithm works as follows:
  \begin{enumerate}
  \item uniformly at random choose a substitution function $\sigma: {X} \to Z_p$, 
  \item compute $f(\tilde{A}[y]|_{\sigma})$ in $t_f(n)$ time,
  \item check whether or not this value is zero and return the result.
  \end{enumerate}

For the time bound note that $t_f(n)$ is obviously $\ge |X|$. For the error bound
 if the degree of $f(\tilde{A}[y])$ is $d$,
  Corollary~\ref{corollary-zippel-schwartz} shows the probability of a false zero  is
$\le d/p 
\le d_f(n)/p$.
\end{proof}

\begin{theorem}
\label{theorem-meta-2}
If the result of $f$ has degree bounded by some polynomial $d_f(n)$ and
 $f(\tilde{A}[y]|_{\sigma})$ is computable by a straight-line
program in $t_f(n)\ge d_f(n)$ time, then
there exists an algorithm that in $O(t_f(n))$ time
checks for each $x \in X$ whether or not 
$ \frac{\partial}{\partial x} f(\tilde{A}[y])$ is symbolically non-zero over $Z_p$.
All $|X|$ returned results are correct with total error probability
$\le d_f(n)|X|/p$.
\end{theorem}

\begin{proof}
  Assume that the Baur-Strassen theorem gives a routine to compute 
$\partial_x f
:=\frac{\partial}{\partial x} 
f(\tilde{A}[y])$
  for all $x\in X$ in $O(t_f(n))$ time.  The algorithm works as follows:

  \begin{enumerate}
  \item uniformly at random choose a substitution function $\sigma: {X} \to Z_p$, 
  \item apply the Baur-Strassen routine to compute 
$\partial_x f|_{\sigma} $
  for all $x\in X$ in $O(t_f(n))$ time,
  \item for each $x\in X$ check whether or not
$\partial_x f |_{\sigma}$ 
is zero and return the result.
  \end{enumerate}

Obviously $\partial_x f$ has degree $\le d_f(n)$. So
  Corollary~\ref{corollary-zippel-schwartz} shows that  for each $x\in X$
the probability of a false zero for
$\partial_x f$ is
$\le d_f(n)/p$.
  Hence the union bound shows the probability of any false zero in $|X|$ results is 
$\le d_f(n)|X|/p$. 
\end{proof}

%

The true goal is to check if a functional value  is symbolically non-zero with high probability.
This is easy to do with some weak assumptions on $f$. 
Specifically assume that any value $f(\alpha)$ in the range of $f$ is a polynomial whose constant coefficients all 
have absolute value at most $C$, for some constant $C$.
Also assume $t_f(n)\ge n+d_f(n)$.

Now consider the setting of Theorem \ref{theorem-meta-1}.
$f(\alpha)$ is a non-zero polynomial iff it is a non-zero polynomial over $Z_p$ for any prime $p>C$.
In Theorem \ref{theorem-meta-1} take $p$ as a prime of size  $\Theta(n\cdot d_f(n))$. We get error probability
$O(1/n)$. 
The time to find $p$ is easily accounted for by the assumption on $t_f(n)$.
So for $f$ as above,  the theorem shows that in $O(t_f(n))$ time we can check if $f(\tilde{A}[y])$ is symbolically non-zero with high probability.

Next consider the setting of Theorem \ref{theorem-meta-2}.
Any value of $\partial_x f$ is a non-zero polynomial iff it is a non-zero polynomial over $Z_p$ for any prime $p>Cd_f(n)$.
In Theorem \ref{theorem-meta-2} take $p$ as a prime of size  $\Theta(n|X|\cdot d_f(n))$. We get error probability
$O(1/n)$. The time is similar. So for $f$ as above  the theorem shows that in $O(t_f(n))$ time we can check if 
$ \frac{\partial}{\partial x} f(\tilde{A}[y])$ is symbolically non-zero 
with high probability.

\newcommand{\cB}{\mathcal{B}}
\newcommand{\distance}{\dist}

\section{Minimum Weight Perfect Matching}
\label{section-matchings}
This section presents an algorithm that,
given an undirected graph with integral
edge weights in $[0,W]$, finds a minimum
weight perfect matching in $\tilde{O}(Wn^\omega)$ time, assuming such matchings exist.

The algorithm works in three phases:
\begin{enumerate}
  \item The first phase uses algebra                      
  to reduce the problem to connected graphs, where
  each edge belongs to some minimum weight perfect matching
(Algorithm \ref
{algorithm-matching-allowed}).
  Moreover for each vertex $v$, we are given the value
  $w(M(v))$ -- the minimum weight of a matching with exactly 2 unmatched vertices,
one of which is $v$
(Algorithm \ref{algorithm-matching-mu}).
This phase uses $\tilde{O}(Wn^\omega)$ time and succeeds with high probability.

  \item The second phase
defines a new weight $w'(uv):=w(uv)+w(M(u))+w(M(v))$
for each edge $uv$. It
performs a simple graph search algorithm on these new edges
to
  obtain a laminar family of blossoms, which is the support of some optimum dual solution
(Algorithm \ref{algorithm-matching-blossoms}).
Each blossom induces a factor critical graph. This phase
is deterministic and uses  $\tilde{O}(n^2)$ time.
  \item The last phase uses a maximum cardinality matching algorithm
  (for unweighted graphs), guided by the structure of the blossoms,
  to obtain a minimum weight perfect matching
(Lemma \ref{lem:dual-to-primal}).
This phase uses
$\tilde{O}(n^\omega)$ time and succeeds with high probability.
\end{enumerate}

To elaborate on the second phase (which in our opinion is the most interesting),
let $A$ be the set of distinct values of the weight function $w'$.
Section~\ref{ssec:dual} proves $|A|=O(n)$.
For $\alpha \in A$ we define a 'threshold graph' $G_\alpha=(V,E')$,
which is an unweighted undirected graph with $E'=\{uv \in E : w'(uv) \le \alpha\}$.
The   nontrivial connected components of all the graphs $G_\alpha$
constitute the blossoms of an optimum dual solution!
(A connected component is {\em nontrivial} if it has more than $1$ and less than $n$ vertices.)
Our proof of this result hinges on showing there exists a special dual solution
(called balanced critical dual solution)
in which it is easy to find the blossoms (Lemma \ref{lem:mu-to-blossoms}).

\begin{figure}[h]
\begin{center}
\subfloat{
  \includegraphics[width=0.33\textwidth]{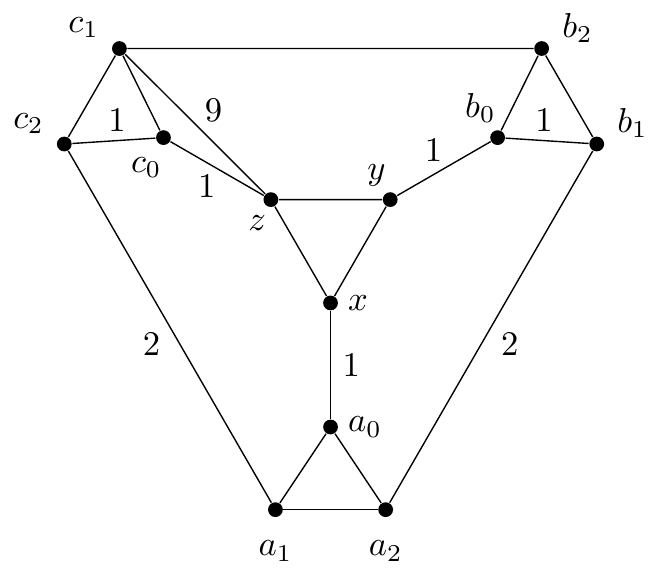}
}
\subfloat{
  \includegraphics[width=0.33\textwidth]{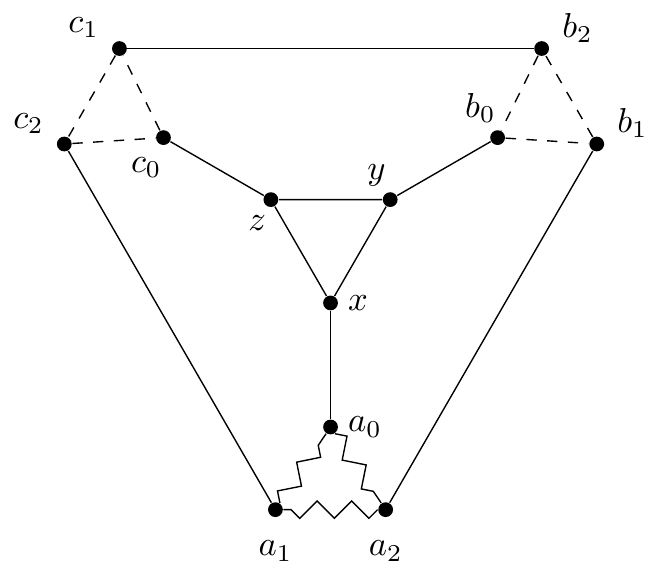}
}
\subfloat{
\hspace{-0.7cm}
  \includegraphics[width=0.33\textwidth]{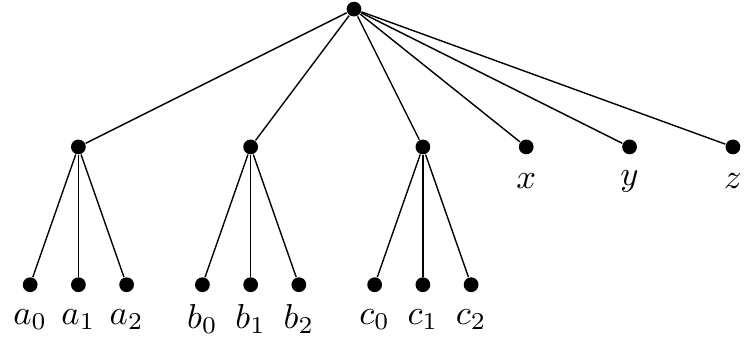}
}
\caption{The matching algorithm: The far left shows an example graph. Edges without a label weigh $0$.
 The minimum weight of a perfect matching is $3$. $w(M(v))$ is 2 for $v \in \{b_2,c_1,x,y,z\}$
  and $1$ otherwise.
  The middle figure shows the allowed edges (edge $c_1z$ was removed) and new edge weights $w'(uv)=w(uv)+w(M(u))+w(M(v))$; edges with $w'(e)=2$ are drawn zigzag, $w'(e)=3$ are dashed, and $w'(e)=4$ are straight.
 The far right shows the laminar family induced by the blossoms found using the threshold graphs 
of Algorithm \ref{algorithm-matching-blossoms}.
}
\label{fig:match-example}
\end{center}
\end{figure}

Figure~\ref{fig:match-example} depicts a sample graph and illustrates the steps for obtaining the laminar
family of blossoms.

In the rest of this section we show how to obtain the set of
{\it allowed edges}, i.e., edges belonging to
at least one minimum weight perfect matching and values $w(M(u))$ (Section~\ref{ssec:matching-tools}).
Next, in Section~\ref{ssec:dual} we present the standard LP
formulation of the problem. We recall and extend
properties of a dual solution, in order to prove the correctness of our simple method of obtaining
blossoms of an optimum dual solution.
Finally, in Section~\ref{ssec:algorithm}, we gather all the theorems
and formally prove correctness and bound the running time of our algorithm.

We would like to note, that the simplicity of our algorithm for contracting the set of blossoms
from the values $w(M(u))$ is due to the fact that the hardness
is hidden in the proof of the purely combinatorial existential lemmas from Section~\ref{ssec:dual}
and in the algorithm for finding unweighted maximum matching problem.


\subsection{Algebraic Tools}
\label{ssec:matching-tools}

Let us define a \emph{symbolic adjacency matrix} of the weighted undirected
graph $G=(V,E,w,W)$ to be the $n \times n$ matrix $\tilde{A}(G)$ such
that
\[
\tilde{A}(G)_{i,j} =
 \left\{
 \begin{array}{rl}
 x_{i,j}y^{w(ij)} & \textrm{if  } {ij} \in E \textrm{ and } i < j, \\
 -x_{j,i}y^{w(ij)} & \textrm{if  } {ij} \in E \textrm{ and } i > j, \\
 0 & \textrm{otherwise,}
\end{array}
\right.
\]
where $x_{i,j}$ are unique variables corresponding to the edges $ij\in E$ of $G$.
Karp, Upfal and Wigderson~\cite{karp-upfal-wigderson-86}
proved that the smallest degree of $y$ in $\det(\tilde{A}(G))$
is twice the weight of a minimum weight perfect matching in $G$.
By using this line of reasoning
together with results of Storjohann and Baur-Strassen,
we show how to obtain the set of edges which appear in
at least one minimum weight perfect matching.

\begin{lemma}
\label{lem:edge-allowed}
An edge $ij \in E$ belongs to
some minimum weight perfect matching
iff
$$\frac{\partial}{\partial x_{i,j}} \term^*_y\left[ \det(\tilde{A}(G))\right] \neq 0\,.$$
\end{lemma}

\begin{proof}
By the definition of a determinant we have:
  \begin{equation}
\label{MatchDetEqn}
    \det (\tilde{A}(G)) =
    \sum_{p \in \Gamma_n} \sgn(p) \prod_{k=1}^{n}
    \tilde{A}(G)_{k,p_k},
  \end{equation}
  where $\Gamma_n$ is the set of $n$-element permutations.
  A permutation $p$ defines a multiset of edges $\mathcal{C}_p =
  \{\{i,p_i\} : 1 \le i \le n \textrm{ and } i\neq p_i\}$. This edge
  multiset corresponds to a cycle cover given by the cycles of $p$.
  Reversing an odd cycle
  in a permutation does not change its sign, but it changes
  the sign of the monomial corresponding to this permutation
  in the determinant.
  Consequently, the polynomial $\det(\tilde{A}(G))$
  contains only monomials corresponding to even-cycle covers of $G$.
  (An {\em even-cycle cover}  has no odd cycles.)
Since each even-cycle cover is easily decomposable into
  two perfect matchings, and doubling a matching
  gives an even-cycle cover, we infer that
  an edge $ij \in E$ belongs to some minimum weight perfect
  matching in $G$ iff it appears in some minimum degree
  monomial in $\det(\tilde{A}(G))$. The lemma follows.
\end{proof}

Note that an even-cycle cover corresponding to a matching -- i.e., every cycle has length 2 -- corresponds
to a unique permutation in \eqref{MatchDetEqn}. Thus the lemma is true in any field $Z_p$.

The combination of Lemma~\ref{lem:edge-allowed} with Theorem~\ref{theorem-meta-2} proves the following corollary.

\begin{corollary}
\label{cor:allowed}
For a weighted undirected graph $G=(V,E,w,W)$
one can compute the set of edges which belong to at
least one minimum weight perfect matching in $\tilde{O}(Wn^\omega)$ running time, with high probability.
\end{corollary}

\begin{algorithm}[H]
\begin{algorithmic}[1]
\State{Generate a random substitution $\sigma:X \to Z_p$ for a prime $p$ of order $\Theta(n^2)$.}
\State{Compute $d= \deg^*_y\left[\det(\tilde{A}(G)|_{\sigma})\right]$ using Storjohann's theorem.}
\State{Let $\partial_x f$ be the routine given by
the Baur-Strassen theorem to compute the matrix of partial derivatives
$\frac{\partial}{\partial x_{i,j}} \term^d_y
\left[\det(\tilde{A}(G))\right]$. }
\State{Generate a random substitution $\sigma:X \to Z_p$ for a prime $p$ of order $\Theta(n^4)$.}
\State{Compute the matrix $\delta=\partial_x f|_{\sigma}$.}
\State{Mark each edge $ij$, where $i<j$, as allowed if $\delta_{i,j} \neq 0$.}
\end{algorithmic}
\caption{Computes the set of allowed edges in the graph $G$.}
\label{algorithm-matching-allowed}
\end{algorithm}

\begin{definition}[$\mathbf{M(uv)}$, $\mathbf{M(u)}$]
For a pair of vertices $u,v \in V$ let $M(uv)$
be a minimum weight perfect matching in $G\setminus \{u,v\}$, i.e.,
$G$ with vertices $u$ and $v$ removed.
Similarly, for a vertex $u$ let $M(u)$ be
a minimum weight almost-perfect matching in the graph $G\setminus \{u\}$.
\end{definition}

Note that $M(u)$ always exists, since we assume the given graph $G$ has a perfect matching.
In contrast $M(uv)$ needn't exist. In that case $M(uv)$ is $\infty$.

\begin{lemma}
  \label{lemma-matching-adjoint}
  Let $G = (V,E,w,W)$ be a weighted undirected graph. Then
  $\deg^*_y(\adj(\tilde{A}(G))_{i,j}) = w(M) + w(M(ij))$.
\end{lemma}

\begin{proof}
  We have $\adj(\tilde{A}(G))_{i,j} = (-1)^{i+j}
  \det(\tilde{A}(G)^{j,i})$. Equivalently, if we take $\tilde{Z}$
  to be the matrix obtained from $\tilde{A}(G)$ by zeroing entries
  of the $j$'th row and the $i$'th column and setting the entry
  $(j,i)$ to 1, then $\adj(\tilde{A}(G))_{i,j} = \det(\tilde{Z})$,
  and so
  \begin{equation}
    \label{equation-lemma-matching-adjoint}
    \adj(\tilde{A}(G))_{i,j} = \sum_{p \in \Gamma_n} \sgn(p)
    \prod_{k=1}^{n} z_{k,p_k}.
  \end{equation}
\noindent  
The permutation $p$ can be
 viewed as a set of directed cycles $\mathcal{C}$ covering $G$, where
  there is a cycle $c$ that contains the edge $(j,i)$. (By definition $(j,i)$
is an edge of  $\mathcal{C}$, even when $(j,i)\notin E(G)$. $(j,j)$ is an example.
The other edges of $\mathcal{C}$ are in $E(G)$.)

We claim that the terms in this adjoint correspond to 
 even-length-cycle covers that  contain a cycle $c$
 through $(j,i)$. In other words cycle covers that contain an odd cycle
make no net contribution to \eqref{equation-lemma-matching-adjoint}.
In proof let $d$ be an odd cycle in $\mathcal{C}$.
First suppose $d\ne c$.
Reversing  its direction changes the sign
  of its contribution to \eqref{equation-lemma-matching-adjoint},
 from antisymmetry of $\tilde{A}(G)$. So such covers do not contribute to the adjoint.
Second suppose $d=c$. 
$G$ has an even
  number of vertices. Thus
$\mathcal{C}$ has an even number of edges, and $c$ cannot be the unique odd cycle.


Now take any $\mathcal{C}$ contributing to \eqref{equation-lemma-matching-adjoint}.
$\mathcal{C}$ decomposes into
 two matchings by 
  taking alternate edges from each (even) cycle.
One matching, say $N$, is a perfect matching of $G$; the other is a 
perfect matching of $G \setminus \{i,j\}$, say $N(ij)$, plus edge $(j,i)$.
We conclude that the adjoint of \eqref{equation-lemma-matching-adjoint}
is 0 if $N$ or $N(ij)$ does not  exist.
This proves the case of the lemma when $M$ or $M(ij)$ does not exist.

Now assume both $M$ and $M(ij)$ exist.
The
  degree of any term in \eqref{equation-lemma-matching-adjoint}
equals the sum of the
  edge weights in $\mathcal{C}$ (note that the weight of the edge $(j,i)$
  is considered as $0$, since $z_{j,i}=1$).
Take any term that has the smallest degree in \eqref{equation-lemma-matching-adjoint}.
Its degree is
\[
  \deg^*_y(\adj(\tilde{A}(G))_{i,j}) = w(N) + w(N(ij)) \ge w(M)+w(M(ij)).
  \]

We will complete the proof by showing
  \[
  \deg^*_y(\adj(\tilde{A}(G))_{i,j}) \le w(M) + w(M(ij)).
  \]
(Obviously the two displayed inequalities show equality holds, so they imply
the lemma.)
The multiset $M \cup M(ij) \cup (j,i)$ gives an even-length-cycle cover
$\mathcal{C}_M$
of $G$ which contains the edge $(j,i)$. Wlog
 the cycle through $(j,i)$ is the only cycle in $\mathcal{C}_M$
(since we can make $M$ and $M(ij)$ identical on any other even cycle).
We claim the monomial corresponding to $\mathcal{C}_M$
occurs exactly once in \eqref{equation-lemma-matching-adjoint}.
This claim shows the monomial does not get cancelled, thus proving the
desired inequality.

To prove the claim,
a variable $x_{gh}$ appearing in any
term comes from the cycle cover edge $(g,h)$ or $(h,g)$. This implies
that two terms with the same variables are cycle covers that differ only in the orientation
of some cycles. 
A cycle of $\mathcal{C}_M$ either has length two or contains $(j,i)$. The former
does not change when it is reversed, and the latter cannot be reversed.
So no other cycle cover corresponds to the monomial of $\mathcal{C}_M$.

Note that since the monomial for $\mathcal{C}_M$ occurs exactly once, the lemma holds
for any field $Z_p$.
\end{proof} 
\noindent
If $M$ or $M(ij)$ does not exist, the expression of the lemma equals $\infty$.
The lemma and the following extension also hold in any field $Z_p$.

\begin{corollary}
\label{cor:mu1}
The vector of values $ w(M) + w(M(i))$, $i\in V$ equals $\deg^*_y
\left(\adj(\tilde{A}(G))b\right) $ for 
$b$ a vector of $n$ indeterminates $b=(b_1,b_2,\ldots,b_n)$.
\end{corollary}

\begin{proof}
Lemma \ref{lemma-matching-adjoint} shows 
$ w(M) + w(M(i))= \min_j \deg^*_y \left(  \adj(\tilde{A}(G))_{i,j}\right)$.
(Note that entries $ \adj(\tilde{A}(G))_{i,j}$ corresponding to the 0 polynomial
cause no problem.)
The $i$th component of the vector $\adj(\tilde{A}(G))b$
is $\sum_j    \adj(\tilde{A}(G))_{i,j}b_j$, and because of the indeterminates $b_j$
no terms cancel when the sum is formed.
 Thus $\deg^*_y
\left(  \adj(\tilde{A}(G))b\right)= w(M) + w(M(i))$.
\end{proof}

This leads to the following algorithm to compute   $w(M(u))$ for all $u \in V(G)$.
Let $B=\{b_1,\ldots,b_n\}$.

\begin{algorithm}[H]
\begin{algorithmic}[1]
\State{Generate a random substitution $\sigma:X \cup B\to Z_p$ for a prime $p$ of order $\Theta(n^3)$.}
\State{Compute $w(M)=\deg^*_y(\det(\tilde{A}(G)|_\sigma))/2$ and $v=\det(\tilde{A}(G)|_\sigma)(\tilde{A}^{-1}(G)b)|_\sigma$ using Storjohann's theorem, where $b=(b_1,\ldots,b_n)$.}
\State{For each $i \in V$ set $w(M(i))=\deg^*_y(v_i)-w(M)$.}
\end{algorithmic}
\caption{Computes the values $w(M(i))$ for all vertices $i$ in the graph $G$.}
\label{algorithm-matching-mu}
\end{algorithm}

To see this algorithm is correct Statement 2 computes $v =
\big(\adj( \tilde{A}(G))b\big)|_\sigma$. So Corollary \ref{cor:mu1} 
shows the algorithm is correct if there are no false zeroes.
A rational expression (like those in $(\tilde{A}^{-1}(G)b)|_\sigma $)
is zero if and only if its 
numerator is zero and its
denominator is nonzero.
So we can apply
the Schwartz-Zippel Lemma to show the final products
have no false zeroes (in their lowest order term).
So each $\deg^*_y(v_i)$ is computed correctly with high probability.

Regarding efficiency consider the $n$ multiplications of degree $nW$
polynomials done to form $v$ in Statement 2. We only use the lowest degree term of each product
(Statement 3). That term comes from  the lowest degree term in  $\det(\tilde{A}(G)|_\sigma$
and 
the lowest degree term in the numerator and the denominator of $(\tilde{A}^{-1}(G)b)|_\sigma $.
So we can find the smallest degree of $y$ that corresponds to $\deg^*_y(v_i)$ using $O(1)$ additions and subtractions, without multiplying polynomials.

We conclude:

\begin{corollary}
\label{cor:mu}
Algorithm \ref{algorithm-matching-mu}
computes  the values $w(M(u))$, for all $u \in V(G)$,
in $\tilde{O}(Wn^\omega)$ time, with high probability.
\end{corollary}

\subsection{Properties of the Dual}
\label{ssec:dual}

We move on to the linear programming formulation of the minimum weight perfect
matching problem given by Edmonds~\cite{e}. An {\em odd set} has odd cardinality; $\Omega$ denotes
the collection of odd subsets of $V$ of cardinality 
$\ge 3$.

\begin{eqnarray}
\min \sum_{e\in E} w(e) x_e && \nonumber\\
x(\delta(v))&=&1, \textrm{ for all } v\in V\nonumber\\
x(\delta(U)) &\ge& 1, \textrm{ for all } U \in \Omega \label{eqn:primal} \\
x_{e} &\ge& 0, \textrm{ for } e \in E\nonumber
\end{eqnarray}

The variables $x_{e}$ indicate when an edge is included in the solution. Here, $\delta(U)$ denotes
all edges $uv \in E$ having $|\{u, v\}\cap U|=1$. We write $\delta(u)$ for $\delta(\{u\})$ and
 $x(F)$ for $\sum_{e\in F} x_e$.

The dual problem has  variables $\pi_v$ for each vertex $v$ and $\pi_U$ for each odd set $U$:\begin{eqnarray}
\max \sum_{v\in V} \pi_v + \sum_{U \in \Omega} \pi_U&&\nonumber\\
\pi_u + \pi_v + \sum_{U\in \Omega,\ uv\in \delta(U)} \pi_U  &\le& w(uv) \textrm{ for all } uv\in E \label{eqn:dual}\\
\pi_U &\ge& 0 \textrm{ for all } U \in \Omega\nonumber
\end{eqnarray}


We say that an edge $e=uv$ is {\it tight} with respect to a dual $\pi$ if 
equality holds in \eqref{eqn:dual}.
A {\em laminar family} is a set system where each pair of sets is either disjoint or one set
contains the other.
Moreover, a graph is {\em factor critical} if after removing each vertex the graph has a perfect matching.
We use existence of the following dual:

\begin{lemma}[Edmonds '65~\cite{e}]
\label{lem:critical}
There exists an optimal dual solution $\pi : V \cup \Omega \rightarrow \mathbb{R}$, such that:
\begin{enumerate}
  \item the set system $\{U \in \Omega : \pi_U > 0\}$ forms a laminar family,
  \item for each $U \in \Omega$ with $\pi_U > 0$, the graph $G[U]$
  with each set of $\{S \in \Omega : S \subset U, \pi_S>0\}$ contracted 
  is factor critical.
\end{enumerate}
\end{lemma}

\begin{definition}[{\bf critical dual, blossom}]
An optimum dual solution satisfying the conditions from Lemma~\ref{lem:critical}
is a {\em critical dual solution}.
A set $U \in \Omega$ such that $\pi_U > 0$ is a {\em blossom}
w.r.t. $\pi$.
\end{definition}

Blossoms of critical dual solutions
have the following useful property
(note the lemma below is weight-oblivious and the only input given to the algorithm
is an undirected unweighted graph, the family of blossoms, and $v$).

\begin{lemma}
\label{lem:blossom-structural}
Consider any critical dual solution and let $U \in \Omega$ be an
arbitrary blossom.
For any vertex $v \in U$ there exists a perfect matching $M(U,v)$
in $G[U \setminus \{v\}]$,
such that for each blossom $U_0 \subseteq U$,
$|M(U,v)\cap\delta(U_0)|$ is 0
if $v \in U_0$ and 1 if $v \not\in U_0$.
Furthermore, given the family of all blossoms and $v$, one can find such a matching
in $\tilde{O}(|U|^\omega)$ running time, with high probability.
\end{lemma}

\begin{proof}
Let $\cB$ be the set of blossoms of $\pi$
properly contained in $U$ ($\cB$ might be empty);
moreover let $\cB_{\max}$ be the set of inclusionwise maximal
sets in $\cB$.
Let $G'$ be the graph $G[U]/\cB_{\max}$
and let $v'$ be a vertex of $G'$
corresponding to $v$. (Here we use the contraction operator -- if $\cal S$ is a family of disjoint vertex sets,
$G/{\cal S}$ denotes the graph $G$ with each set of $\cal S$ contracted to a single vertex.)

Initially let $M(U,v) \subseteq E$ be a perfect matching in $G'\setminus v'$. It
exists since $G'$ is factor critical.
For each blossom $U_0 \in \cB_{\max}$,
recursively find a perfect matching $M_0$ in the graph $G[U_0 \setminus x]$,
where $x$ is the single vertex of the intersection of $U_0$ and $V(M(U,v)) \cup \{v\}$. 
  Add the edges of $M_0$ to  $M(U,v)$.

By construction the final matching $M(U,v)$ satisfies conditions from the lemma.
For the time bound note that the
 laminarity of $\cB$ implies the total number of vertices in all graphs
constructed by the above procedure is $O(n)$.
The algorithms of \cite{ms04,Sankowski05p,Harvey06} find
a perfect matching on an arbitrary graph of $n$ vertices  in time
$\tilde{O}(n^\omega)$, with high probability. Hence our recursive procedure
runs in total time $\tilde{O}(|U|^\omega)$.
\end{proof}

Complementary slackness gives the following observation.

\begin{observation}
\label{obs:slackness}
For any optimum  dual solution:
\begin{enumerate}
  \item[(a)] a set $U \in \Omega$ with $\pi_U > 0$
has exactly one edge of $\delta(U)$ in
  any minimum weight perfect matching; 
  \item[(b)] an edge belonging  to any minimum weight perfect matching is tight.
\end{enumerate}
\end{observation}

\begin{lemma}
\label{lem:dual-to-primal}
Given a weighted undirected graph $G=(V,E,w,W)$ where each edge is allowed,
and the set of blossoms $\cB$ of some critical dual solution,
one can find a minimum weight perfect matching in $\tilde{O}(n^\omega)$ time,
with high probability.
\end{lemma}

\begin{proof}
Let $\cB_{\max} \subseteq \cB$ be the set of inclusionwise maximal blossoms.
Let graph $G'=G/\cB_{\max}$. $G'$ has a perfect matching $M_0$ (since
Observation~\ref{obs:slackness}(a) shows any minimum weight perfect matching in $G$
contains a subset of edges forming a perfect matching in $G'$).

We extend $M_0$ to a perfect matching in $G$ by
considering the blossoms of $U\in \cB_{\max}$ one by one.
Let $v_U$ be
the unique vertex of $U$ that is matched by $M_0$.
Add the matching $M(U,v_U)$ 
of  Lemma~\ref{lem:blossom-structural}
to  $M_0$.
The final set $M_0 $
is a perfect matching for $G$.

The time for this  procedure is $\tilde{O}(n^\omega)$.
This follows exactly as in  Lemma~\ref{lem:blossom-structural}, since the total number of
vertices in all graphs considered is $O(n)$.

Finally, we prove that $M$ is a minimum weight perfect matching in $G$
by showing that for each blossom $U \in \cB$ the set $M$ contains
exactly one edge of $\delta(U)$.
Consider a maximal blossom $U \in \cB_{\max}$.
When finding a perfect matching in $G'$ we have added exactly a single
edge of $\delta(U)$ to the set $M$.
Moreover each edge of $M(U',v_{U'})$, for any maximal blossom $U' \in \cB_{\max}$,
is contained in $U'$, and therefore the set $M$ contains exactly one edge of $\delta(U)$.
Now let us consider a blossom $U \in \cB \setminus \cB_{\max}$,
and let $U' \in \cB_{\max}$ be a maximal blossom containing $U$.
If in the first phase (finding a perfect matching in $G'$)
we added no edge from $\delta(U)$ to the set $M$, then
by Lemma~\ref{lem:blossom-structural}
in the set $M(U',v_{U'})$ there is exactly one edge of $\delta(U)$,
whereas for each other maximal blossom $U'' \in \cB_{\max}, U'' \neq U'$,
in the set $M(U'',v_{U''})$ there is no edge of $\delta(U)$.
If, however, in the first phase we added an edge from $\delta(U)$ to the set $M$,
then by the choice of $v_U'$, which is the endpoint of the edge of $\delta(U) \cap M$,
no other edge of $\delta(U)$ is added to the set $M$.

Since all the edges are allowed, by Observation~\ref{obs:slackness}(b)
the sum of values of edges of $M$ is equal to the cost
of the critical dual solution, which proves that $M$ is a
minimum weight perfect matching in the graph $G$.
Note that our algorithm does not need exact values of the dual nor
even weights of edges, since its input is only graph $G$ and set $\cB$.
\end{proof}

A
critical dual solution gives rise to a weighted tree in a natural way:

\begin{definition}[{\bf dual tree}]
Let $\pi : \Omega \cup V \rightarrow \mathbb{R}$ be a critical dual solution,
with $\cB$ the set of its blossoms.
The {\em dual tree} $T(\pi)$ is a rooted tree on  nodes $\{V\} \cup \cB \cup V$,
where $V$ is the root, vertices of $V$ are leaves, blossoms of $\cB$
are internal nodes and the parent-child relation in $T(\pi)$ is naturally defined inclusionwise.
The weight of the
edge from a node $t \in \cB \cup V$
to its parent is $\pi_t$. The height of the tree $H(T(\pi))$
is the weight of a longest path from the root to some leaf.
\end{definition}

In this definition note that the last edge of a path defining $H(T(\pi))$ may have negative length.
For a tree $T$ with weighted edges and  two nodes $u,v$, $\distance_T(u,v)$ denotes
the weight of the path between $u$ and $v$.
The following simple lemma provides a basic tool.

\begin{lemma}
\label{lem:matching-path}
If $\pi$ is a critical dual solution for a weighted graph $G=(V,E,w,W)$,
any allowed edge $uv$ satisfies $w(uv) = \distance_{T(\pi)}(u,v)$.
\end{lemma}

\begin{proof}
Since $uv$ is tight (Observation~\ref{obs:slackness}(b)),
$w(uv)=\pi_u + \pi_v + \sum_{U \in \Omega,\ uv\in \delta(U)} \pi_U$.
The right-hand side gives $\distance_{T(\pi)}(u,v)$ for two reasons:
The edges of $T(\pi)$ incident to leaves are weighted with
the singleton values of $\pi$.
A blossom $B$ of $\pi$ contains exactly one endpoint of the edge $uv$
if and only if the path between $u$ and $v$ in $T(\pi)$ contains
the edge between $B$ and its parent.
\end{proof}

The next steps of our development (Lemmas \ref{lem:balanced-paths}--\ref{HeightLemma})
can be derived using an appropriate version of Edmonds' weighted matching algorithm
(e.g., \cite{Schrijver}). Here we will use a structural approach, based on
the following properties of  allowed edges given by
Lov\'asz and Plummer.

\begin{lemma}[\cite{lp}, Lemma 5.2.1 and Theorem 5.2.2]
\label{lem:canonical-partition}
Let $G=(V,E)$ be an undirected connected graph 
where each edge belongs to some perfect matching.
Define a binary relation $R\subseteq V \times V$ 
by $(u,v) \in R$ if and only if 
$G\setminus \{u,v\}$ has {\em no} perfect matching.
Then
\begin{itemize}
  \item $R$ is an equivalence relation; 
  \item each equivalence class of $R$ is an independent set;
  \item for each equivalence class $S$ of $R$,  the graph $G\setminus S$ has exactly $|S|$ connected components, 
  each of which is factor critical.
\end{itemize}
\end{lemma}

We will use a special type of critical dual solution that
we call "balanced".

\begin{definition}[{\bf balanced critical dual}]
\label{def:balanced}
Let $\pi : \Omega \cup V \rightarrow \mathbb{R}$
be a critical dual solution, and let $G'$ be
the graph $G$ with each blossom of $\pi$ contracted.
$\pi$ is a {\em balanced critical dual solution} if there are 
two distinct vertices $u,v \in V$ such that $\distance_{T(\pi)}(u,V)=\distance_{T(\pi)}(v,V)=H(T(\pi))$
and further,
$G' \setminus \{u',v'\}$  has a perfect matching for 
$u',v'$ the (distinct) vertices of $G'$ corresponding to $u,v$, respectively.
\end{definition}

Before
proving that balanced critical dual solutions exist,
 we give a lemma showing why they are useful. In particular they show
how 
the $M(v)$ values relate to $T(\pi)$.
Let $M(G)$ be a minimum weight perfect matching in $G$.

\begin{lemma}
\label{lem:balanced-paths}
Let $G=(V,E,w,W)$ 
be an undirected connected graph 
with every edge in some minimum weight perfect matching.
Let $\pi$ be a balanced critical dual solution for $G$.
For any vertex $z \in V$, 
a minimum weight almost perfect matching in $G \setminus z$ weighs
$w(M(G)) - H(T(\pi)) - \distance_{T(\pi)}(z,V)$.
\end{lemma}

\begin{proof}
 Any almost perfect matching in $G \setminus z$ weighs at least
$w(M(G)) - H(T(\pi)) - \distance_{T(\pi)}(z,V)$. In proof
let $M_1$ be an arbitrary perfect matching in 
 $G \setminus \{x,z\}$ 
for any $x\in V$.
For any blossom $U$ of $\pi$ such that $x,z\not\in U$,
$|M_1\cap \delta(U)|\ge 1$.
Together with 
\eqref{eqn:dual} this gives
 $w(M_1)\ge 
\sum_{w\in V-x,z} \pi_w +\sum_{x,z\notin U} \pi_U$.
The right-hand side equals $w(M(G))-\pi_x-\pi_z-\sum_{\{x,z\}\cap  U\ne \emptyset} 
\pi_U$, by strong duality. Since every $\pi_U$ is nonnegative this quantity is at least
$w(M(G)) - \distance_{T(\pi)}(z,V) - \distance_{T(\pi)}(x,V)$. The definition of $H(T(\pi))$ shows
the last quantity is at least $w(M(G)) - \distance_{T(\pi)}(z,V) - H(T(\pi))$ as desired. 

We complete the proof by constructing an almost perfect matching in $G \setminus z$ of weight
$w(M(G)) - H(T(\pi)) - \distance_{T(\pi)}(z,V)$.
Take $G',u,v,u',v'$ as in Definition~\ref{def:balanced}.
Moreover let $z'$ be the vertex of $G'$ corresponding to $z$.
$G'$ is connected, with every edge in a perfect matching, so it satisfies the
hypothesis of  Lemma \ref{lem:canonical-partition}.
Definition~\ref{def:balanced} shows that
$u' \not \hskip-4pt R v'$. So $z'$ is not equivalent to
at least of $u$ and $v$. 
W.l.o.g. assume that $u' \not \hskip-4pt R z'$.
Thus $G' \setminus \{u',z'\}$ has a perfect matching $M_0$.

Next, consider each inclusionwise maximal blossom $U$ of $\pi$ one by one.
Let $x\in U$ be the unique vertex of $U$
in the set $V(M_0)\cup \{u,z\}$.
Add to
$M_0$ the edges of the matching $M(U,x)$
 guaranteed by Lemma~\ref{lem:blossom-structural}.

Clearly $M_0$ is a perfect matching in $G \setminus \{u,z\}$.
For each blossom $U$ of $\pi$,   $|M_0\cap\delta(U)|$ is 1
if $u,z \not\in U$, and 0 if $u$ or $z$ belongs to $U$.
Blossoms of the latter type are  those  in  the path
from $u$ to $V$ or $z$ to $V$ in $T(\pi)$. 
These two paths have disjoint edge sets, since
 $u'\ne z'$.
We get $w(M_0)=w(M(G)) -  \distance_{T(\pi)}(u,V) -\distance_{T(\pi)}(z,V) $,
since every edge of $M_0$ is allowed, i.e., tight.
Since $\distance_{T(\pi)}(u,V)=
H(T(\pi))$  this is the desired weight.  
\end{proof}

We prove that balanced critical duals
exist in two steps.
The first step shows a simpler property for  critical duals  actually makes them balanced. The second step shows duals with this property exist. 

\begin{lemma}
Let $G=(V,E,w,W)$ 
be an undirected connected graph 
with every edge in some minimum weight perfect matching.
A critical dual $\pi_0$ is balanced if it has minimum height (i.e.,
 $H(T(\pi_0))$ is no larger than the height of any other critical dual).
\end{lemma}

\begin{proof}
Assume for the purpose of contradiction  that $\pi_0$ is not a balanced critical dual.
For any vertex $v\in V$ let $h_v$ denote its height in $\pi_0$, $h_v=\distance_{T(\pi_0)}(v,V)$.
Let
$u$ be the vertex of $G$ with the greatest height $h_u$.
Let $G'$ be the graph $G$ with inclusionwise maximal blossoms of $\pi_0$
contracted. Let $R$ be the equivalence relation 
of
Lemma~\ref{lem:canonical-partition} for $G'$, and
$S_1,\ldots,S_k$ its equivalence classes. Let $u$ belong to vertex $u'$ of $G'$ and let
$u' \in S_1$. 

We will define a dual function $\pi_1$. An element of $S_1$
is either a maximal blossom of $\pi_0$ or a vertex of $V$ not in any blossom;
let  $s_i$, $1\le i\le |S_1|$, be the $i$th of these blossoms and vertices.
Lemma \ref{lem:canonical-partition} shows $G'\setminus S_1$
has $|S_1|$ connected components; let
$B_i$, $1\le i\le |S_1|$, be the set of vertices of $G$ contracted onto  the $i$-th connected component of $G'\setminus S_1$.
Define $\pi_1 : \Omega \cup V \rightarrow \mathbb{R}$ to be identical to $\pi_0$ except
\[\pi_1(x)=
\begin{cases}
\pi_0(x) - \epsilon&x =s_i,\ 1\le i\le |S_1|\\
\pi_0(x)+\epsilon&x=B_i,\ 1\le i\le |S_1|.
\end{cases}
\]
(Note that if $B_i$ consists of more than one vertex in $G'$ then we are creating a new blossom.)
Let $\epsilon$ be any positive real no larger than the smallest
value of $\pi_0(s_i)$ for a blossom $s_i$. This ensures $\pi_1$ is nonnegative on blossoms.

\begin{figure}[h]
\begin{center}
\includegraphics{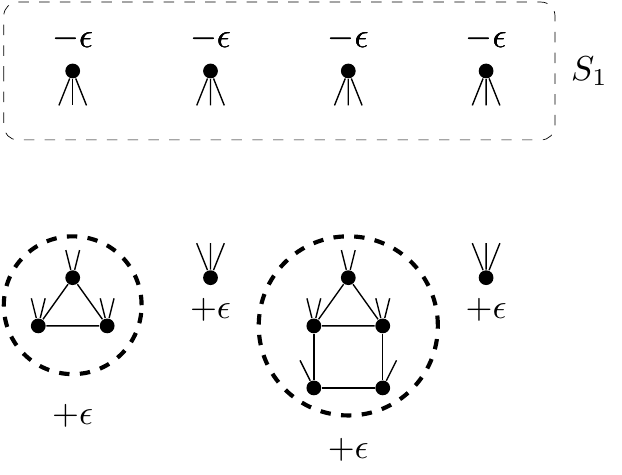}
\label{fig:matching}
\caption{Graph $G'$ and  the modification of the duals.}
\end{center}
\end{figure}

Let us verify that $\pi_1$ is a critical dual.
First, observe that each edge of $G$ remains tight in $\pi_1$:
Nothing changes for an edge that has both ends in the same set of
$S_1$ or some $B_i$. The remaining possibility is an edge
between $S_1$ and some $B_i$ (no edge joins 2 $B_i$ sets or 2 sets of $S_1$,
the latter by independence of
$S_1$). For such edges
we have added and subtracted $\epsilon$ in the 
left-hand side of \ref{eqn:dual}, so it remains tight.
Next observe that the blossoms of $\pi_1$ form a laminar family. Lemma~\ref{lem:canonical-partition} 
shows the sets $B_i$ induce factor critical graphs. Finally $\pi_1$ is an optimum dual,
since its  objective as $\pi_0$. Thus $\pi_1$
is a critical dual.

Taking $\epsilon $ small enough makes 
$\pi_1$ 
a critical dual with smaller height than $\pi_0$, the desired contradiction.
To see this take
 any
vertex $v \in V$, and let $v'$ be
the vertex of $G'$ that  $v$ is contracted onto.
If $v'\in S_1$, the height of $v$ decreases as long as
$\epsilon$ is positive.
Suppose $v' \notin S_1$.
Lemma~\ref{lem:canonical-partition} shows
$\pi_0$ would be balanced if  $h_v=h_u$.
Thus $h_v < h_u$. Choose
$\epsilon$ small enough so that every such $v$
has 
$\distance_{T(\pi_1)}(v)=\distance_{T(\pi_0)}(v)+\epsilon \le h_u - \epsilon$.
Thus $\pi_1$ has smaller height than $\pi_0$.
\end{proof}

\begin{lemma}
\label{HeightLemma}
Let $G=(V,E,w,W)$ 
be an undirected connected graph 
with every edge in some minimum weight perfect matching.
There is a critical dual $\pi_0$ that has the smallest height $H(T(\pi_0))$.
\end{lemma}

\begin{proof}
Lemma~\ref{lem:critical} shows a critical dual $\pi$ exists. 
There are a finite number of
laminar families on $V$, i.e., a finite number of trees $T(\pi)$. 
So it suffices to show that there is a smallest height among all critical duals $\pi$
with the same tree  $T=T(\pi)$. 

We begin by showing that for every blossom $U$, there is a unique value for $\pi_x$, where
 $x$ is any vertex of $U$ or  any blossom
properly contained in
$U$. We argue inductively, so assume this holds for every blossom properly contained in $U$.
For any edge $uv$ we break the left-hand side of \eqref{eqn:dual}
into the contributions from $u$ and from $v$, by defining
\[\pi_{u,v}=\pi_u + \sum_{U\in \Omega,\ u=\{u,v\}\cap U}\pi_U \]
and symmetrically for $\pi_{v,u}$. So the left-hand side of  \eqref{eqn:dual}
is $\pi_{u,v}+\pi_{v,u}$.

 Take any edge $uv$ joining two vertices  $u,v\in U$.
$uv$ is on an odd cycle $C$ contained in $U$. 
($U$ is factor critical, so let $M_u$ ($M_v$) be a perfect matching on $U-u$ ($U-v$) respectively.
The symmetric difference $M_u\oplus M_v$ contains an even-length path from $u$ to $v$.)
Each edge of $C$ is tight. So for every edge $xy$ in $C$, the values of  $\pi_{x,y}$
 and $\pi_{y,x}$ are uniquely determined. If $\pi_{x,y}$ does not have any contributions
from blossoms properly contained in $U$ then $\pi_x=\pi_{x,y}$ has been  uniquely determined.
If $\pi_{x,y}$ has a contribution $\pi_W$ from a blossom $W$ that is a maximal blossom
properly contained in $U$ then $\pi_W$ has been uniquely determined.
This follows since the other $\pi$ values contributing to $\pi_{x,y}$ have been determined by induction.
(Note that $\pi_W$ has also been uniquely determined from the other edge of $C\cap \delta(W)$.)
If neither of these conditions apply to $\pi_{x,y}$ then all its $\pi$-values have been determined by induction.
Since any vertex $u\in U$ is on an edge $uv$ in $U$, this completes the inductive argument.

Next consider any edge $uv$ not contained in a blossom of $T$. 
The previous argument
shows exactly one term in the quantity  $\pi_{u,v}$ is still undetermined.
If $uv$ is in an odd cycle $C$ the previous argument shows that term is uniquely determined.
Contract all such odd cycles as well as all blossoms of $T$. We get a bipartite graph $G'$.
It contains at least one edge.
Let $S$ be a spanning tree of $G'$. Choose a value $p_0$ for the unknown term $p$
at the root of $S$, that comes from a valid critical dual for $T$.  Suppose we increase $p$.
If this increases $H(T)$,  every value of $p$ larger than $p_0$ gives larger height.
Suppose this decreases $H(T)$.
All the other unknown $\pi$-values are uniquely determined from tightness of the edges of $S$.
Also every  edge of $G'$ not in $S$ remains tight by bipartiteness. There is a maximum value $\overline p$
such that every value $p>\overline p$ either makes the $\pi$-values invalid 
(because some $\pi_U$, $U\in \Omega$ becomes
negative) or increases the height (since $p$ contributes to the height of the root vertex).
Similarly there is a minimum value $\underline p$ for $p$. We conclude there is a unique
smallest height for 
a critical dual for $T$ -- it occurs when $p$ is equal to either $\overline p$ or $\underline p$.
\end{proof}


As already mentioned, the last two lemmas show any undirected connected graph 
$G=(V,E,w,W)$ with all edges allowed
has a balanced critical dual. We can now reach our final goal.

\begin{lemma}
\label{lem:mu-to-blossoms}
Let $G=(V,E,w,W)$ be
a weighted undirected connected graph 
where every edge  is allowed.
Given all  values $w(M(v))$ for $v \in V$,
the  blossoms  of
a balanced critical dual solution can be found in $\tilde{O}(n^2)$
time.
\end{lemma}

\begin{proof}
Let  $\pi$
be a balanced critical dual solution.
By Lemma~\ref{lem:balanced-paths} for each leaf node $v \in V$ of $T(\pi)$,
 $w(M(v)) = w(M(G)) - H(T(\pi)) - \distance_{T(\pi)}(z,V)$.
Define new edge weights $w':E \rightarrow \mathbb{Z}$
as $w'(uv) = w(M(u)) + w(M(v)) + w(uv)$.
Consider any $uv\in E$.
Since $uv$ is tight, $w(uv) = \distance_{T(\pi)}(u,v)$.
Define a quantity 
$c$ that is independent of $uv$, 
$c=2(w(M(G))-H(T(\pi)))$.
Then
\begin{eqnarray}
w'(uv) & = & w(uv) + w(M(u)) + w(M(v)) \nonumber\\
    & = & \distance_{T(\pi)}(u,v) + 2w(M(G)) - 2H(T(\pi)) - \distance_{T(\pi)}(u,V) - \distance_{T(\pi)}(v,V)
\nonumber\\
    & = & 2(w(M(G))-H(T(\pi))) - 2\distance_{T(\pi)}(\lca(u,v),V)\nonumber\\
&=&  c-2\distance_{T(\pi)}(\lca(u,v),V)\,.\label{wEqn}
\end{eqnarray}

Let $B=\lca(u,v)$. So $B$ is
the inclusionwise minimal blossom of $\pi$ containing both $u$ and $v$,
or
if no such blossom exists, $B$ is
the root $V$ of the tree $T(\pi)$. 
For any edge $uv$ let $B_{uv} \subseteq V$ be  the
the set of vertices  reachable from $u$ or $v$ by a path of edges $e$ satisfying
$w'(e) \le w'(uv)$.

\newenvironment{claim_num}[1]{\noindent {\bf Claim #1.}\em}{}
\newenvironment{claim_num_proof}[1]{\noindent{\em Proof of Claim #1.}}{\hfill\hfill$\diamond$\bigskip}

\newenvironment{no_num_claim}{\bigskip\noindent {\bf Claim.}\em\bigskip}{}
\newenvironment{no_num_claim_proof}{\noindent{\em Proof of Claim. }}{\hfill\hfill$\Diamond$\bigskip}

\begin{no_num_claim}
For any edge $uv$,  $B_{uv}=B$. 
\end{no_num_claim}

\begin{no_num_claim_proof}
Let $F \subseteq E$ be the set of edges of a spanning tree of $G[B]$ ($G[B]$ is connected
since either $B=V$ or $G[B]$ is factor critical).
Since any edge $ab\in F$ is contained in $B$, 
the node $\lca(a,b)$ descends from $\lca(u,v)$ in $T(\pi)$.
Thus the path from $\lca(a,b)$ to $\lca(u,v)$ in $T$ has nonnegative weight.
This implies $w'(ab) \le w'(uv)$  by \eqref{wEqn}.
Thus $B \subseteq B_{uv}$.

For the opposite inclusion, consider any edge $ab$ with $a\in B$ and 
$w'(ab) \le w'(uv)$.
Since every blossom has a strictly positive $\pi$-value,  \eqref{wEqn} implies
$b\in B$. Now an easy\
induction shows any path from $u$ or $v$, with every  edge $e$ having $w'(e)\le w'(uv)$,
has every vertex in $B$. Thus $B_{uv} \subseteq B$.
\end{no_num_claim_proof}

Any blossom $B$ of $\pi$ has an edge $uv$ with  $B$ the minimal blossom containing $u$ and $v$
(by laminarity and connectedness of $B$).
So the claim of the lemma amounts to constructing
all the sets $B_{uv}$. This is 
done in $\tilde{O}(n^2)$ time by  Algorithm~\ref{algorithm-matching-blossoms} below.
\end{proof}

\begin{algorithm}[H]
\begin{algorithmic}[1]
\State For each edge $uv$ set $w'(uv)=w(uv)+w(M(u))+w(M(v))$.
\State{Let $A$ be the set of all different values $w'(uv)$. Let $\mathcal{B}=\emptyset$.}
\For{each $\alpha \in A$, in increasing order,}
\State{Let $\mathcal{C}$ be the set of connected components
  of the graph $(V,\{uv : uv \in E, w'(uv) \le \alpha\})$.}
\State Add the nontrivial components of $\mathcal{C}$ to $\mathcal{B}$.
\EndFor
\State \Return $\mathcal{B}$.
\end{algorithmic}
\caption{Given all the values $w(M(u))$, finds the blossoms of a balanced critical dual in the graph $G$ where all edges are allowed.}
\label{algorithm-matching-blossoms}
\end{algorithm}

\subsection{The Final Algorithm}
\label{ssec:algorithm}

\begin{theorem}
Let $G=(V,E,w,W)$ be a weighted undirected graph containing a perfect matching.
A minimum weight perfect matching in $G$ can be computed
$\tilde{O}(Wn^\omega)$ time, with high probability.
\end{theorem}

\begin{proof}
First, using Corollary~\ref{cor:allowed}, we can remove all the edges of $G$ which
are not allowed.
Clearly, we can consider each connected component of $G$ separately,
hence w.l.o.g. we assume that $G$ is connected.
Next, compute all the values $w(M(u))$ for each $u \in V$
using Corollary~\ref{cor:mu}.
Having all the values $w(M(u))$ by Lemma~\ref{lem:mu-to-blossoms}
we can find the set of blossoms $\cB$ of a balanced critical dual solution
and consequently by Lemma~\ref{lem:dual-to-primal}
we can find a minimum weight perfect matching in $G$.
\end{proof}

The full version of this paper shows how the matching algorithm can be made Las Vegas.

In some applications the second smallest perfect matching is of interest
(e.g., Section \ref{section-undirected-negative}). Its weight 
is easily found, as follows.
As discussed in the proof of Lemma~\ref{lemma-matching-adjoint}, the terms in the determinant of
$\det(\tilde{A}(G))$ correspond to even-cycle covers in the graph $G$. Each such cycle can be decomposed
into two perfect matchings. As we already observed the smallest degree term in $y$ in $\det(\tilde{A}(G))$
corresponds to taking twice the minimum weight perfect matching in $G$. The next smallest
term gives the following.

\begin{corollary}
\label{corollary-det-second}
Let $G = (V,E,w,W)$ be a weighted undirected graph.
The degree in $y$ of a second smallest monomial  of $\det(\tilde{A}(G))$ is equal to
the weight of a minimum weight perfect matching $M^*$ plus the weight of a
second smallest perfect matching $M'$. In particular, the weight of
a second smallest perfect matching can be found in $\tilde{O}(W n^{\omega})$ time, with high probability.
\end{corollary}

\newlength{\lw}
\setlength{\lw}{\textwidth}
\addtolength{\lw}{-20pt}

\begin{proof}
  Letting $\tau(M^*)$  denote the term  corresponding to $M^*$
in $\det(\tilde{A}(G))$,\hfill\break
\hbox to \lw{\hfill$w(M')=\term^*_y\left[ \det(\tilde{A}(G))-\tau(M^*)\right].$\hfill}
\end{proof}

\section{Diameter and Radius}
\label{section-diameter}

In this section we consider the problem of computing the diameter and radius
of a directed graph without negative weight cycles. By bidirecting the
edges this result can be applied to undirected graphs with nonnegative edge weights.

\def\eccentricity{\mathit{eccentricity}}

We start with definitions of the quantities of interest, plus equivalent definitions
that we use to compute these quantities. To motivate
the latter note that
it seems difficult to compute a given set of distances $S$ directly. Instead
we show how to check if, for an arbitrary value $c$,
 all the distances in $S$ are $\le c$. For a graph $G$
and a vertex $i$,
\[\begin{array}{lclcl}
eccentricity(i)&=& \max \{\dist_G(i,j):j\in V\} &=& \min \{c: (\forall j) (c\ge \dist(i,j))\},\\
radius(G)&=& \min \{eccentricty(i):i\in V\}&=& \min \{c: (\exists i)( \forall j) (c\ge \dist(i,j))\},\\
diameter(G)&=& \max \{eccentricty(i):i\in V\}&=& \min \{c: (\forall i,j)( c\ge \dist(i,j))\}.
\end{array}
\]

We use the following theorem proven in~\cite{Sankowski05}\footnote{In \cite{Sankowski05} the graph was defined to contain self-loops whereas here we add self-loops in the equation by taking $\tilde{A}(\overrightarrow{G})+I$.}:
\begin{lemma}
  \label{lemmaPathsEsa} Let $\overrightarrow{G}$ be a directed weighted graph without negative weight cycles.
  The weight of the shortest path in $G$ from $i$ to $j$ is given by
 $\dist_G(i,j) = \deg^*_y\left(\adj\left(\tilde{A}(\overrightarrow{G})+I\right)_{i,j}\right)$.
Moreover, all non-zero terms in $\det\left(\tilde{A}(\overrightarrow{G})\right)_{i,j}$ are
  non-zero over any finite field $\mathcal{Z}_p$.
\end{lemma}

In order to be able to use the above lemma we first need to apply the following observation.

\begin{corollary}
\label{corollary-workaround-4}
Let $c$ be arbitrary number from $[-nW,\ldots,nW]$. There exists $d \le c$ such that
$\term^d_y\left[\adj\left(\tilde{A}(\overrightarrow{G})+I\right)_{i,j}\right] \neq 0$
if and only if
$\term^c_y\left[\adj\left(\tilde{A}(\overrightarrow{G})+I\right)_{i,j}\cdot\left(\sum_{i=0}^{2nW}y^i\right)\right] \neq 0$.
This continues to hold in any finite field.
\end{corollary}
\begin{proof}
Multiplication of a polynomial by $\sum_{i=0}^{2nW}y^i$ means that we add all lower degree terms to a term of a given degree. Hence,
if some degree term $d$ was non-zero then all higher degree terms become non-zero
(assuming no terms get cancelled).
Since $0\le c-d\le 2nW$ a degree $d$ term creates a corresponding degree $c$ term in the product.

Finally observe that no term drops out because of cancellations\footnote{We shall apply this principle
to other matrices.}: In 
the matrix $\tilde{A}(\overrightarrow{G})+I$, every nonzero entry has the form
$xy^{w(x)}$, where $x$ is the indeterminate $\ne y$ (or 1 in diagonal entries) and $w$ is a function.
So every term in $\det(\tilde{A}(\overrightarrow{G})+I)$ has the exponent of $y$ functionally dependent
on the remaining variables. This holds for an entry of the adjoint too.
 Thus
a term $(\prod x)y^c$ created in the multiplication of the corollary comes from exactly one
 term $(\prod x)y^d$ in the adjoint
(i.e., $d=\sum w(x)$). So there are no cancellations, in ordinary arithmetic or in
$Z_p$.
\end{proof}

To find the diameter, we use
this Corollary to perform a binary search for the lowest $c$ such that
for all $i,j \in V$, 
\[\term^c_y\left[\adj\left(\tilde{A}(\overrightarrow{G})+I\right)_{i,j}\cdot\left(\sum_{i=0}^{2nW}y^i\right)\right] \neq 0.\]
Clearly this
$c$ is the diameter. Similarly a binary search for the lowest $c$ where the displayed condition
holds for some $i$ with every $j$ gives the radius.

The main problem left is how to check whether $\term^d_y\left[\adj\left(\tilde{A}(\overrightarrow{G})+I\right)_{i,j}\cdot\left(\sum_{i=0}^{2nW}y^i\right)\right] \neq 0$. We will show how to obtain $\adj$ as a partial derivative of $\det$. We define $\tilde{Z}$ to be {\it a fully symbolic matrix}
of size $n \times n$ as $\tilde{Z}_{i,j} = z_{i,j}$, where $z_{i,j}$ are unique variables for all $i, j \in [1,\ldots,n]$. We define
$\sigma_z$ to be an evaluation that assigns $0$ to all $z_{i,j}$. Now we are ready to prove the following lemma.

\begin{lemma}
\label{lemma-partial-adjoint}
\[
\term^d_y\left[\adj\left(\tilde{A}(\overrightarrow{G})+I\right)_{i,j}\left(\sum_{i=0}^{2nW}y^i\right)\right] =
\frac{\partial}{\partial z_{j,i}} \term^d_y\left[\det\left(\tilde{A}(\overrightarrow{G})+I+\tilde{Z}\right)\left(\sum_{i=0}^{2nW}y^i\right)\right]\Bigg|_{\sigma_z}.
\]
\end{lemma}

\begin{proof}
Observe that for the fully symbolic $n\times n$ matrix $\tilde{Z}$ and any $n\times n$ matrix $\tilde{A}$
not involving any variable $z_{i,j}$,
\begin{equation}
\label{PartialsForAdj}
\frac{\partial}{\partial z_{j,i}} \det(\tilde{A}+\tilde{Z})|_{\sigma_z}=
\adj(\tilde{A})_{i,j}.
\end{equation}
Thus
\[
\frac{\partial}{\partial z_{j,i}} \term^d_y\left[\det\left(\tilde{A}(\overrightarrow{G})+I+\tilde{Z}\right)\!\left(\sum_{i=0}^{2nW}y^i\right)\right]\Bigg|_{\sigma_z}\!\!\!\! =
\term^d_y\left[\frac{\partial}{\partial z_{j,i}}\det\left(\tilde{A}(\overrightarrow{G})+I+\tilde{Z}\right)\Bigg|_{\sigma_z} \!\!\! \left(\sum_{i=0}^{2nW}y^i\right)\right]\! 
\]
\[
=\term^d_y\left[\adj\left(\tilde{A}(\overrightarrow{G})+I+\tilde{Z}\right)_{i,j}\Big|_{\sigma_z} \left(\sum_{i=0}^{2nW}y^i\right)\right]
=\term^d_y\left[\adj\left(\tilde{A}(\overrightarrow{G})+I\right)_{i,j} \left(\sum_{i=0}^{2nW}y^i\right)\right].
\]
\end{proof}


Joining the above results together with Theorem~\ref{theorem-meta-2} and Corollary~\ref{corollary-workaround-4} we get the following algorithm.

\begin{algorithm}[H]
\begin{algorithmic}[1]
\State{Let $\partial_z f$ be the routine given by
the Baur-Strassen theorem to compute the matrix of partial derivatives
$\frac{\partial}{\partial z_{i,j}} 
\term^{c+nW}_y
\left[ 
 \left(\sum_{i=0}^{2nW}y^i\right)
 \det\left(  (\tilde{A}(\overrightarrow{G})+I+\tilde{Z})  y^{W}\right) 
\right]$. }
\State{Generate a random substitution 
$\sigma:X \to Z_p$ for a prime $p$ of order $\Theta(n^4)$.
Extend it to
$\sigma:X \cup Z\to Z_p$ by setting
$\sigma|_Z=\sigma_z$.}
\State{Compute the matrix  $\delta=\partial_z f|_{\sigma}$.}
\State{Return true if $\delta_{i,j}$ is non-zero for all $i,j\in V$.}
\end{algorithmic}
\caption{Checks whether diameter of the directed graph $\overrightarrow{G}$ is $\le c$.}
\label{algorithm-diameter}
\end{algorithm}

Note that the polynomial multiplication in Step 1 need only compute the coefficient of $y^{c+nW}$
and so only uses time $O(nW)$.
Using binary search with the above algorithm, and using a similar algorithm for radius, we obtain:

\begin{theorem}
  \label{theorem-diameter}
  Let $\overrightarrow{G} = (V,E,w,W)$ be a weighted directed graph without negative weight cycles.
  The diameter and radius of $G$ can be computed in $\tilde{O}(W n^{\omega})$ time, with high probability.
\end{theorem}

\section{Shortest Cycles in Undirected Graphs}
\label{section-cycles-u}
Let $G=(V,E,w,W)$ a weighted undirected graph. In this section we consider only the problem of computing 
shortest cycles when all edges have non-negative weights. The more general case is solved in the next section, and 
uses the ideas introduced here. Let us define a \emph{symbolic adjacency matrix} of the weighted undirected
graph $G$ to be the be the symbolic matrix polynomial $\tilde{A}(G)$ equal to $\tilde{A}(\overrightarrow{G})$, where $\overrightarrow{G}$
is the bidirection of $G$.

\begin{lemma}
\label{lemmaDeterminantU}
  Let $G$ be an undirected weighted graph with no negative edges, and 
  $d \in [0,nW]$. Some $uv \in E$ has
 $\left[x_{u,v}\frac{\partial}{\partial x_{u,v}}-x_{v,u}\frac{\partial}{\partial x_{v,u}}\right]\term^d_y\left[\det(\tilde{A}(G)+I)-1\right]\neq 0$
  iff some cycle packing of weight $d$ contains a component that is an oriented
  cycle through $uv$ in $G$.
  Moreover, for $p\ge 2$, all non-zero terms in
the above expression 
  are
  non-zero over the  finite field ${Z}_p$.
\end{lemma}
\begin{proof}
As argued in the proof of Lemma~\ref{lemmaDeterminant} $\det(\tilde{A}(\overrightarrow{G})+I)-1$ contains only terms that correspond to cycle packings. Moreover, the degree of each non-zero term is equal to the total weight of the cycles in the packing. However, because $\tilde{A}(G)$ was constructed using bidirected graph there might be terms containing both antiparallel edges, that correspond to bidirected edges in the undirected graph.

  Next we show that a cycle packing $\mathcal{C}$ in $G$
  contributes to the expression of the lemma iff
  it contains exactly one of the variables $x_{u,v}$ and $x_{v,u}$,
  and hence contains a simple cycle passing through $uv$.
  Moreover we show that in such a case the contribution of $\mathcal{C}$
  is a product of variables corresponding to $\mathcal{C}$ 
  and hence the contribution of $\mathcal{C}$ is not cancelled out by a 
  different cycle packing.
  If $\mathcal{C}$ does not contain $x_{u,v}$ or $x_{v,u}$,
   clearly it has zero contribution. This leaves two possibilities:

\paragraph{{\bf Case 1.} $\mathcal{C}$ contains a cycle $u,v,u$} 
Since
$\left[x_{u,v}\frac{\partial}{\partial x_{u,v}}-x_{v,u}\frac{\partial}{\partial x_{v,u}}\right] x_{u,v}x_{v,u} =
x_{u,v}x_{v,u}-x_{v,u}x_{u,v} = 0,$
$\mathcal{C}$'s term makes no contribution.

\paragraph{{\bf Case 2.} 
{$\mathcal{C}$ contains a simple cycle $C$ containing $uv$}}
The corresponding term contains exactly one of $x_{u,v}$ and $x_{v,u}$
say $x_{u,v}$.
We have
$\left[x_{u,v}\frac{\partial}{\partial x_{u,v}}-x_{v,u}\frac{\partial}{\partial x_{v,u}}\right] x_{u,v} = x_{u,v}$.
Hence, the derivative for this term is nonzero and is equal
to the sign of permutation multiplied by the product of the variables 
of the oriented edges of $\mathcal{C}$.
\end{proof}

Similarly as in undirected graphs we say that edge $e$ is
{\it allowed} if and only if it belongs to some simple shortest cycle $C$ in $G$. The above proof
actually gives us a way to find allowed edges as well.

\begin{corollary}
\label{corollary-allowed-U}
Let $d$ be smallest number in $[1,nW]$ such that there exists
an edge $uv \in E$ such that
 $\left[x_{u,v}\frac{\partial}{\partial x_{u,v}}-x_{v,u}\frac{\partial}{\partial x_{v,u}}\right]\term^d_y\left[\det(\tilde{A}(G)+I)-1\right]\neq 0$.
Then an edge $uv \in E$ is allowed if and only if when
$\left[x_{u,v}\frac{\partial}{\partial x_{u,v}}-x_{v,u}\frac{\partial}{\partial x_{v,u}}\right]\term^d_y\left[\det(\tilde{A}(G)+I)-1\right]\neq 0$.
\end{corollary}

%

\paragraph{Computing the Weight of the Shortest Cycle}
\label{section-weight-cycle-U}

Using Lemma~\ref{lemmaDeterminantU} we will devise an algorithm that will be
able to check whether there exists a simple cycle in $G$ of length shorter or equal to $c$. In order to
do it we need the observation similar to Corollary~\ref{corollary-workaround-4}.

\begin{corollary}
\label{corollary-workaround-3}
Let $c$ be arbitrary number from $[1,nW]$. There exists $d \le c$ such that
$$\displaystyle \left[x_{u,v}\frac{\partial}{\partial x_{u,v}}-x_{v,u}\frac{\partial}{\partial x_{v,u}}\right]\term^d_y\left[\det(\tilde{A}(G)+I)-1\right] \neq 0$$
if and only if
$\left[x_{u,v}\frac{\partial}{\partial x_{u,v}}-x_{v,u}\frac{\partial}{\partial x_{v,u}}\right]\term^c_y\left[(\det(\tilde{A}(G)+I)-1)\cdot(\sum_{i=0}^{nW}y^i)\right] \neq 0$.
\end{corollary}

Using the above observation we can construct the following algorithm.

\begin{algorithm}[H]
\begin{algorithmic}[1]
\State{Let $\partial_x f$ be the routine given by
the Baur-Strassen theorem to compute the matrix of partial derivatives
$\frac{\partial}{\partial x_{u,v}} 
\term^{c}_y
\left[
(\sum_{i=0}^{nW}y^i)
\left(
\det(\tilde{A}(G)+I)-1\right)\right]$.}
\State{Generate a random substitution 
$\sigma:X \to Z_p$ for a prime $p$ of order $\Theta(n^4)$.}
\State{Compute the matrix  $\delta=\partial_x f|_{\sigma}$.}
\State{Compute the matrix $\delta'$ with $\delta'_{u,v}=\left[x_{u,v}\delta_{u,v}-x_{v,u}\delta_{v,u}\right]\big|_{\sigma}$.}
\State{Return true if $\delta'$ has a non-zero entry.}
\end{algorithmic}
\caption{Checks whether the shortest cycle in undirected graph $G$ has weight $\le c$.}
\label{algorithm-cycles-U}
\end{algorithm}

The correctness of the above algorithm is implied by both Theorem~\ref{theorem-meta-1}
and Theorem~\ref{theorem-meta-2}. Using binary search with it we obtain.

\begin{theorem}
  \label{theorem-shortest-cycle-weight-U}
  Let $G = (V,E,w,W)$ be a weighted undirected graph without negative weight edges.
  The weight of the shortest simple cycle in $G$ can be computed in $\tilde{O}(W
  n^{\omega})$ time, with high probability.
\end{theorem}

\paragraph{Finding the Shortest Cycle}
\label{section-finding-shortest-cycle-U}
After showing how to compute the shortest cycle length it remains to show
how to find the cycle itself. We essentially can use the same approach as we used for
directed graphs in Section~\ref{section-finding-shortest-cycle}.

\begin{algorithm}[H]
\begin{algorithmic}[1]
\State{Let $c^*$ be the weight of the shortest cycle computed using Theorem~\ref{theorem-shortest-cycle-weight-U}.}
\State{Let $\delta'$ be the matrix computed by Algorithm~\ref{algorithm-cycles-U}
for $c=c^*$.}
\State{Take any edge $uv$ such that $\delta_{uv}\neq 0$.}
\State{Compute the shortest path $p_{v,u}$ from $v$ to $u$ in $G \setminus \{uv\}$ using Dijkstra.}
\State{Return the cycle formed by $uv$ and $p_{v,u}$.}
\end{algorithmic}
\caption{Computes the shortest cycle in undirected graph $G$.}
\label{algorithm-cycles-U-2}
\end{algorithm}

\begin{theorem}
  \label{theorem-cycles-U}
  Let $G = (V,E,w,W)$ be a weighted undirected graph without negative weight edges.
  The shortest simple cycle in $G$ can be found in $\tilde{O}(W
  n^{\omega})$ time, with high probability.
\end{theorem}

\section{Undirected Graphs with Negative Weights}
\label{section-undirected-negative}
This section gives algorithms for shortest cycle and  diameter 
in undirected graphs with possibly negative edges but no negative
weight cycles. To accomplish this 
we need to combine the ideas from Sections~\ref{section-diameter} and~\ref{section-cycles-u} 
 with our results for  weighted matching. We will recast the results for the diameter and shortest cycles into
the language of matchings. Hence (unlike Section \ref{section-cycles-u}) throughout this section
the symbolic adjacency matrix $\tilde{A}({G})$ of an undirected graph $G$
is defined as in
Section \ref{ssec:matching-tools}.

\paragraph{Diameter}
Let $G=(V,E,w,W)$ be an undirected graph with negative weights allowed, and let $E^{-}$ be the set of edges with negative weights.
We will define a graph $\ddot{G}$ that models paths in $G$ by almost perfect matchings.
We believe the construction is essentially due to   Edmonds~\cite{Edmonds67}.

Define
the {\it split graph} $\ddot{G}=(\ddot{V},\ddot{E},\ddot{w},W)$ where
\[
\ddot{V} = \{v_1, v_2: v \in V\} \cup \{e_1,e_2: e \in E^{-}\},
\]
\begin{eqnarray}
\ddot{E} = \{v_1 v_2: v\in V\} &\cup& \{u_1v_2, u_2v_1, u_1v_1, u_2v_2 : uv\in E \setminus E^{-}\} \nonumber \\
&\cup& \{u_1 e_1, u_2e_1, e_1 e_2, v_1 e_2, v_2 e_2 : e=uv \in E^{-}, u<v\},\nonumber
\end{eqnarray}
\[
\ddot{w}(u_i v_j) =
 \left\{
\begin{array}{rl}
 w(uv) & \textrm{if  } uv \in E\setminus E^{-}, \\
 w(e) & \textrm{if  } u_i=e_1 \textrm{ and } v_j\neq e_2 \textrm{ and } e \in E^{-},\\
 0 & \textrm{otherwise.}
\end{array}
\right.
\]

\begin{figure}[h]
\begin{center}
  \includegraphics{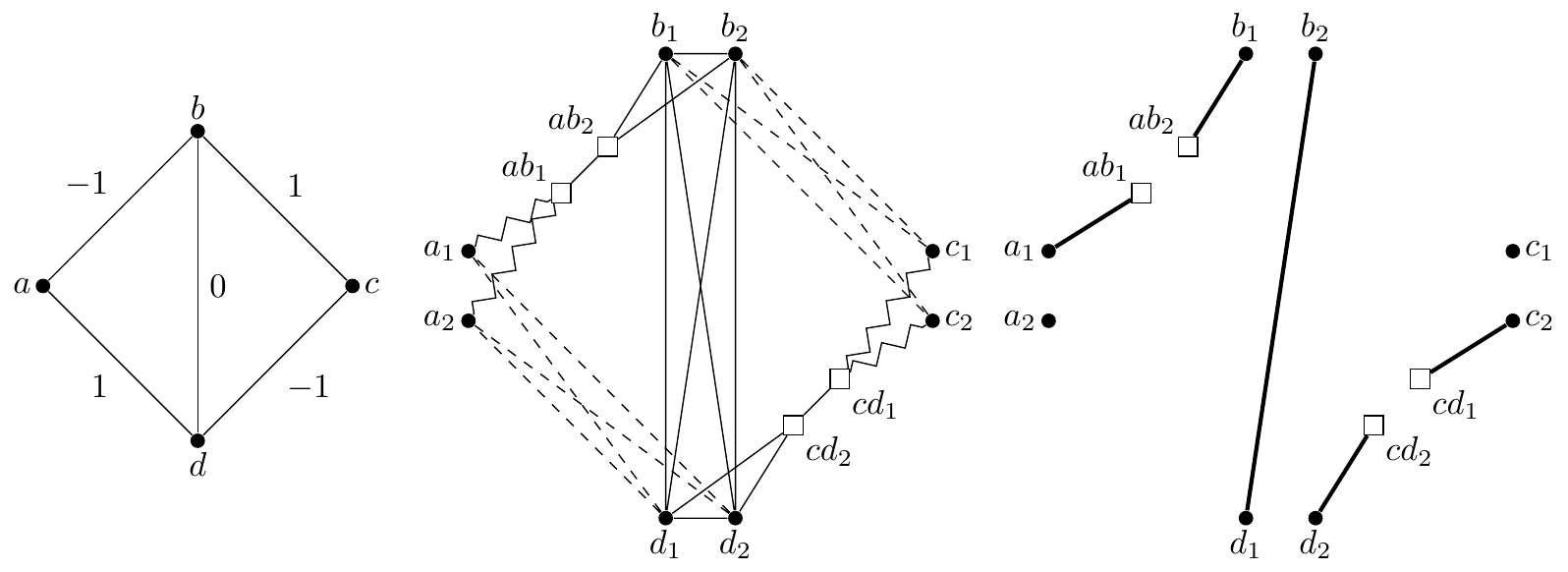}
\caption{An undirected graph $G$ and its graph $\ddot{G}$.
  In $\ddot{G}$ zigzag edges weigh $-1$, dashed edges weigh $1$
    and the remaining edges weigh $0$.
  Vertices corresponding to negative edges of $G$ are white squares.
  The far right  shows a matching $M(a_2c_1)$ of weight $-2$, which corresponds
  to a shortest path between $a$ and $c$.
}
\label{fig:ddot-example}
\end{center}
\end{figure}

Note how a length-two path in $G$, say $a,b,c$ with $w(ab)\ge 0>w(bc)$ and $e=bc$, corresponds to
a matching  in $\ddot{G}$ such as $a_1b_1, b_2e_1,e_2c_1$, having the same total weight.
Fig.\ref{fig:ddot-example} gives a complete example.

An important property is that we can assume
$\ddot{n}=|\ddot{V}| \le 4n$.  This follows since we can assume
 $|E^{-}| < n$, as otherwise the set of negative edges contains a cycle. 

We will consider minimum weight perfect matchings in $\ddot{G}$. 
To use our algebraic tools we should eliminate negative weights by
setting $w'(e):=\ddot{w}(e)+W$. Obviously this
 increases the weight of all perfect matchings by $\ddot{n}W/2$ and so doesn't change the minimum
perfect matching.
But
to keep things simple in the following we keep $\ddot{G}$ as defined above, with $\ddot{w}$ possibly negative.

The following observation is essentially given in~\cite{Ahuja93} in Chapter~12.7 (for a larger
version of our graph).

\begin{lemma}
\label{lemma-lawler}
Let $u,v\in V$, let $M$ be the minimum weight perfect matching, and let $M(u_2v_1)$ be the minimum weight almost perfect matching in $\ddot{G}$
that does not match $v_1$ nor $u_2$. If $G$ does not contain negative weight cycles then $\ddot{w}(M)=0$ and
the shortest path weight from $u$ to $v$ in $G$ is equal to $\ddot{w}(M(u_2v_1))$. \end{lemma}

Note also that it is easy to detect a negative cycle in $G$ -- it corresponds to a perfect matching
in $\ddot{G}$ with negative weight.

By Lemma~\ref{lemma-matching-adjoint} we know that $\deg^*_y(\adj(\tilde{A}(\ddot{G}))_{u_2,v_1}) = \ddot{w}(M) + \ddot{w}(M(u_2v_1))$. Thus
$\dist_G(u,v)=\deg^*_y(\adj(\tilde{A}(\ddot{G}))_{u_2,v_1})$, i.e., just as in
Lemma~\ref{lemmaPathsEsa},  $\adj(\tilde{A}(\ddot{G}))$ encodes the distances in $G$. Now we proceed
exactly as in Algorithm~\ref{algorithm-diameter}.

\begin{algorithm}[H]
\begin{algorithmic}[1]
\State{Let $\partial_z f$ be the routine given by
the Baur-Strassen theorem to compute the matrix of partial derivatives
$\frac{\partial}{\partial z_{u_2,v_1}} 
\term^{c+\ddot{n}W}_y
\left[ 
 \left(\sum_{i=0}^{2nW}y^i\right)
 \det\left(  (\tilde{A}(\ddot{G})+\tilde{Z})  y^{W}\right) 
\right]\hskip-4pt , u,v \in V$. }
\State{Generate a random substitution 
$\sigma:X \to Z_p$ for a prime $p$ of order $\Theta(n^4)$.
Extend it to
$\sigma:X \cup Z\to Z_p$ by setting
$\sigma|_Z=\sigma_z$.}
\State{Compute the matrix  $\delta=\partial_z f|_{\sigma}$.}
\State{Return true if $\delta_{u_2,v_1}$ is non-zero for all $u,v\in V$.}
\end{algorithmic}
\caption{Checks whether diameter of the undirected graph $G$ with negative weights is $\le c$.}
\label{algorithm-diameter-U}
\end{algorithm}

Note that $\tilde{A}(\ddot{G})+\tilde{Z}$ is not skew-symmetric but we still get the adjoint by 
\eqref{PartialsForAdj}.
Similarly to  check if the radius is $\le c$,  Step 4 returns true if 
some row of $\delta$ consists entirely of nonzeroes.
Again using binary search we obtain.

\begin{theorem}
  \label{theorem-diameter-UN}
  Let $G = (V,E,w,W)$ be a weighted undirected graph without negative weight cycles.
  The diameter and radius of $G$ can be computed in $\tilde{O}(W n^{\omega})$ time, with high probability.
\end{theorem}

\paragraph{Shortest Cycles}
Recalling Corollary \ref{corollary-det-second},
it might appear that
a second smallest perfect matching in $\ddot{G}$ corresponds to a shortest cycle in $G$. But
this is not true, because of cycles of length two (e.g., $u_1v_1, v_2u_2$ or
$u_2v_1, v_2u_1$).
These can be handled
as in Section~\ref{section-cycles-u}, by antisymmetric derivatives $x_{u,v}\frac{\partial}{\partial x_{u,v}}-x_{v,u}\frac{\partial}{\partial x_{v,u}}$. 
For the next lemma note that in the absence of negative cycles, any cycle contains
an edge of nonnegative weight.

\begin{lemma}
\label{lemmaDeterminantUN}
 Let $G$ be an undirected weighted graph with no negative  cycle. For any edge
$uv \in E\setminus E^-$,
a shortest cycle through $uv$ weighs
$\term^*_y\left[x_{u_1,v_2}\frac{\partial}{\partial x_{u_1,v_2}}-x_{u_2,v_1}\frac{\partial}{\partial x_{u_2,v_1}}\right]\left(\det(\tilde{A}(\ddot{G}))\right)$.
 This continues to hold in any field $Z_p$,
$p\ge 2$.
\end{lemma}

\begin{proof}
Let $\delta=\left[x_{u_1,v_2}\frac{\partial}{\partial x_{u_1,v_2}}-x_{u_2,v_1}\frac{\partial}{\partial x_{u_2,v_1}}\right]\left(\det(\tilde{A}(\ddot{G}))\right)$.
Observe that in general for any variables $x,y$ and any integers $i,j$,
\begin{equation}
\label{PartialEq}
\left[x\frac{\partial}{\partial x}-y\frac{\partial}{\partial y}\right] x^i y^j =
(i-j) x^i y^j.
\end{equation}
\noindent
Hence the terms in $\delta$ are a subset of those in $\det(\tilde{A}(\ddot{G}))$, i.e., the effect of the
differentiation operator is just to change the multiplicity of some terms, perhaps zeroing them or causing
other cancellations.

Let $C$ be a shortest cycle through $uv$. It gives an even-cycle cover $M^* \cup N$ in
$\ddot{G}$ with
weight $w(C)$,
where 
$M^*$ is
the minimum perfect matching of $\ddot{G}$, $M^*=\{v_1v_2, e_1e_2: v\in V, e \in E^-\}$,
and
 $N$ is the perfect matching of $\ddot{G}$
containing $u_1v_2$ but not $ u_2v_1$, plus representatives of the other edges of
$C$, plus edges $x_1x_2$ for vertices or negative edges $x\notin C$.
Let $\tau$ be the monomial corresponding to this cover
(e.g., $\tau$ has the term $y^{w(C)}$). \eqref{PartialEq} 
(applied to $x^1_{u_1,v_2} x_{u_2,v_1}^0$)
shows $\tau$ is a term contributing to $\delta$.
In fact $\tau$ is the only such term in $\delta$ involving its monomial. This again follows from
\eqref{PartialEq} and the preliminary observation (it is easy to see
$\tau$ is the unique edge cover for its monomial, i.e., none of its cycles can be reversed). We conclude $\term^*_y (\delta )\le w(C)$.

We complete the proof by showing $\term^*_y (\delta) \ge w(C)$. 
First observe that a perfect matching $M$ on $\ddot{G}$ 
that contains $u_1v_2$ but not $ u_2v_1$ weighs  at least $w(C)$.
To prove this
imagine contracting each edge $x_1x_2$ of $\ddot{G}$; call the resulting vertex $x$. Every vertex
now has degree 2 in $M$. 
We will compute the weight of $M$ by examining its edges in the contracted graph.
An edge
$x_1x_2\in M$ becomes a loop at $x$ of weight 0.
An edge $xy\in E\setminus E^-$ with
$x_1y_1, x_2y_2\in M$ or $x_1y_2, x_2y_1 \in M$ becomes 2 copies of $xy$, both
with nonnegative weight. 
The other edges of $M$ form cycles in the contracted graph.  Each cycle is a cycle in $G$
and so has nonnegative weight.
One of the cycles contains edge $uv$, so it weighs at least
$w(C)$. Hence $\ddot{w}(M)\ge w(C)$.

Consider an even-cycle cover $\mathcal{C}$ that contributes to $\delta$.
\eqref{PartialEq} shows $i\ne j$, i.e., $u_1v_2$ and $u_2v_1$ occur with different multiplicities in $\mathcal{C}$. 
The possibilities for $\{i,j\}$ are $\{0,1\}$,  $\{0,2\}$,  and $\{1,2\}$.
$\mathcal{C}$ decomposes into 2 perfect matchings.  In all three cases
one of the matchings of $\mathcal{C}$ contains exactly 1 of the edges $u_1v_2$, $u_2v_1$.
That matching weighs at least $w(C)$. The other matching has nonnegative weight,
so $\mathcal{C}$ weighs at least $w(C)$. In other words $\term^*_y (\delta) \ge w(C)$.
\end{proof}

Using this lemma with the scheme of Algorithm~\ref{algorithm-cycles-U} gives the following.

\begin{algorithm}[H]
\begin{algorithmic}[1]
\State{For $(i,j)=(1,2), (2,1)$, let $\partial_x f^{i,j}$ be the routine given by
the Baur-Strassen theorem to compute the matrix of partial derivatives
$\frac{\partial}{\partial x_{u_i,v_j}} 
\term^{c+\ddot{n}W}_y
\left[
(\sum_{i=0}^{\ddot{n}W}y^i)
\left(
\det(\tilde{A}(\ddot{G})y^W)\right)\right]$.}
\State{Generate a random substitution 
$\sigma:X \to Z_p$ for a prime $p$ of order $\Theta(n^4)$.}
\State{For $(i,j)=(1,2),(2,1)$ compute the matrix  $\delta^{i,j}=\partial_x f^{i,j}|_{\sigma}$.}
\State{Compute the matrix $\delta'$ with $\delta'_{u,v}=\left[x_{u_1,v_2}\delta^{1,2}_{u,v}-x_{u_2,v_1}\delta^{2,1}_{u,v}\right]\Big|_{\sigma}$.}
\State{Return true if $\delta'$ has a non-zero entry.}
\end{algorithmic}
\caption{Checks whether a shortest cycle in undirected graph $G$ with negative weights has weight $\le c$.}
\label{algorithm-cycles-UN}
\end{algorithm}

Again a binary search gives

\begin{theorem}
  \label{theorem-shortest-cycle-weight-UN}
  Let $G = (V,E,w,W)$ be a weighted undirected graph without negative weight cycles.
  The weight of a shortest  cycle can be computed in $\tilde{O}(W
  n^{\omega})$ time, with high probability.
\end{theorem}

Algorithm \ref{algorithm-cycles-UN} with $c$ equal to the
shortest cycle weight gives  an edge $uv$ with
$\delta'_{u,v}\ne 0$, i.e., $uv$ is on a shortest cycle.
So a minimum weight perfect matching on $\ddot{G}-u_1,v_2$
corresponds to a shortest cycle.
Hence
we can state

\begin{theorem}
  \label{theorem-cycles-UN}
  Let $G = (V,E,w,W)$ be a weighted undirected graph without negative weight cycles.
  A shortest  cycle in $G$ can be found in $\tilde{O}(W n^{\omega})$ time, with high probability.
The same holds for a shortest $st$-path, for any given vertices $s,t$.
\end{theorem}

\section{Vertices Lying on Short Cycles}
\label{section-cycles-passing}

For undirected graphs 
we only need to change the output of our algorithms:
For Algorithm~\ref{algorithm-cycles-U} 
(when there are no negative edges)
or Algorithm~\ref{algorithm-cycles-UN}
(in the general case)
we find the
set of vertices lying on cycles of length $\le c$
by changing the last 
step so that it 
returns $\{v\in V: \exists_{vu\in E}\  \delta'_{vu}\neq 0\}$.

For directed graphs we apply the $\sum_{i=0}^{2nW}y^i$ multiplication technique
to Algorithm~\ref{algorithm-cycles-2}:

\begin{algorithm}[H]
\begin{algorithmic}[1]
\State{Let $\partial_x f$ be the routine given by
the Baur-Strassen theorem to compute the matrix  of partial derivatives
$\frac{\partial}{\partial x_{u,v}} \term^{c+nW}_y\left[(\sum_{i=0}^{2nW}y^i)
\left(
\det((\tilde{A}(\overrightarrow{G})+I) y^W)-y^{nW}\right)\right]$.}
\State{Generate a random substitution $\sigma:X \to Z_p$ for a prime $p$ of order $\Theta(n^4)$.}
\State{Compute the matrix $\delta=\partial_x f|_{\sigma}$.}
\State{Return $\{v\in V: \exists_{(v,u)\in E}\  \delta_{(v,u)}\neq 0\}$.}
\end{algorithmic}
\caption{Computes the set of vertices lying on cycles of length $\le c$ in directed graphs.}
\end{algorithm}

Thus we obtain:

\begin{theorem}
  \label{theorem-cycles-passing}
  Let $G$ be a weighted directed or undirected graph with integral weights in $[-W,W]$
and no negative cycle.
For any $c $
  the set of vertices lying on cycles of length $\le c$ can be computed in $\tilde{O}(W
  n^{\omega})$ time, with high probability.
\end{theorem} 


\bibliographystyle{abbrv}
\bibliography{weighted}

\end{document}